\documentclass{aa}

\usepackage[english]{babel}
\usepackage[ansinew]{inputenc}
\usepackage{amssymb}
\usepackage{graphicx,graphics}
\usepackage{natbib}

\newcommand{\prob}{\textrm{prob}}

\newcommand{\diag}{\textrm{diag}}

\newcommand{\SSR}{\textrm{SSR}}

\title{Soft clustering analysis of galaxy morphologies: A worked example with SDSS}

\author{Ren\'e Andrae\inst{1} \and Peter Melchior\inst{2} \and Matthias Bartelmann\inst{2}}
\date{Submitted to A\&A on Feb 01 2010}
\institute{Max-Planck-Institut f\"ur Astronomie, K\"onigstuhl 17, 69117 Heidelberg, Germany\\ e-mail: andrae@mpia-hd.mpg.de \and Institut f\"ur Theoretische Astrophysik, Zentrum f\"ur Astronomie, Albert-Ueberle-Str. 2, 69120 Heidelberg, Germany}

\titlerunning{Soft clustering analysis of galaxy morphologies}
\authorrunning{Ren\'e Andrae et al. (2010)}

\abstract{The huge and still rapidly growing amount of galaxies in modern sky surveys raises the need of an automated and objective classification method. Unsupervised learning algorithms are of particular interest, since they discover classes automatically.}
{We briefly discuss the pitfalls of oversimplified classification methods and outline an alternative approach called ''clustering analysis''.}
{We categorise different classification methods according to their capabilities. Based on this categorisation, we present a probabilistic classification algorithm that automatically detects the optimal classes preferred by the data. We explore the reliability of this algorithm in systematic tests. Using a small sample of bright galaxies from the SDSS, we demonstrate the performance of this algorithm in practice. We are able to disentangle the problems of classification and parametrisation of galaxy morphologies in this case.}
{We give physical arguments that a probabilistic classification scheme is necessary. The algorithm we present produces reasonable morphological classes and object-to-class assignments without any prior assumptions.}
{There are sophisticated automated classification algorithms that meet all necessary requirements, but a lot of work is still needed on the interpretation of the results.}


\begin{document}

\maketitle

\section{Introduction}

Classification of objects is typically the first step towards scientific understanding, since it brings order to a previously unorganised set of observational data and provides standardised terms to describe objects. These standardised terms are usually qualitative, but they can also be quantitative which makes them accessible for mathematical analysis. A famous example of a successful classification from the field of astrophysics is the Hertzsprung-Russell diagram, where stars exhibit distinct groups in the colour-magnitude diagram that represent their different evolutionary stages. For the same reason, galaxy classification is an important conceptual step towards understanding physical properties, formation and evolution scenarios of galaxies.

With the advent of modern sky surveys containing millions (e.g. SDSS, COSMOS, PanSTARRS, GAMA) or even billions (e.g. LSST) of galaxies, the classification of these galaxies is becoming more and more problematic. The vast amount of data excludes the hitherto common practice of visual classification and clearly calls for an automated classification scheme that is more efficient and more objective. In this work, we present an algorithm for automated and probabilistic classification, where the classes are discovered automatically, too. However, the intention of this work is not to come up with ''yet another morphological classification scheme'', but rather to demonstrate of how it could be done alternatively to the standard practice of classification in astrophysics. Besides, we are unable to present a full solution to the problem of morphological galaxy classification, since there is still no accepted method for parametrising arbitrary galaxy morphologies \citep[cf.][]{Andrae2010b}. In addition to the lack of convincing classification schemes, this is why many experts are very sceptical about the subject of classifying galaxy morphologies as a whole. As parametrisation of galaxy spectra is more reliable, spectral classifications have become more accepted.\\
\\*
In the remaining part of this introduction, we first give an overview about modern automated classification methods and work out a categorisation of these methods. We describe our parametrisation of galaxy morphologies using shapelets \citep{Refregier2003} in Sect. \ref{sect:shapelets}. In Sect. \ref{sect:soft_clustering_algorithm} we present the algorithm we are using, which has been introduced before by \citet{Yu2005} in the field of pattern recognition. We extensively investigate the reliability of this classification algorithm in Sect. \ref{sect:systematic_tests}. Such a study has not been undertaken by \citet{Yu2005}. In Sect. \ref{sect:worked_example_SDSS} we present a worked example with a small sample of 1,520 bright galaxies from the SDSS. The objects in this sample are selected such that no practical problems with parametrisation arise, as we want to disentangle the problems of classification and parametrisation as much as possible. The aim of this worked example is \textit{not} to do science with the resulting classes or data-to-class assignments, but to demonstrate that such an algorithm indeed produces reasonable results. We conclude in Sect. \ref{sect:conclusions}.

\subsection{Overview about classification methods}

In Table \ref{tab:summary_classification_algorithms} we give an overview of different classification methods and some example algorithms. The two criteria for this categorisation are:
\begin{enumerate}
\item Is the data-to-class assignment probabilistic (soft) or not (hard)?
\item Are the classes specified a priori (classification) or discovered automatically (clustering)?
\end{enumerate}
Not all algorithms fit into this categorisation, namely those that do not directly assign classes to objects (e.g. self-organising maps).

\begin{table}
\begin{tabular}{ccc}
\hline\hline
Type & Classification & Clustering \\
\hline
Hard & nearest neighbour,     & $K$-means, \\
     & Fisher's linear        & spectral clustering, \\
     & discriminant analysis  & kernel PCA \\
\hline
Soft & na\"ive Bayes,         & Gaussian mixture \\
     & linear/quadratic       & models \\
     & discriminant analysis, &  \\
     & neural networks        &  \\
\hline
\end{tabular}
\caption{Overview of different classification and clustering algorithms with examples.\newline Soft (probabilistic) algorithms are always model-based, whereas hard algorithms are not necessarily. Soft algorithms can always be turned into hard algorithms, but not vice-versa. The list of example algorithms is not complete.}
\label{tab:summary_classification_algorithms}
\end{table}

The algorithm we are going to present is a soft algorithm, i.e. the data-to-class assignment is probabilistic (cf. next section). The reason is that in the case of galaxy morphologies, it is obvious that the classes will \textit{not} be clearly separable. We rather expect the galaxies to be more or less homogeneously distributed in some parameter space, with the classes appearing as local overdensities and exhibiting potentially strong overlap. As we demonstrate in Sect. \ref{sect:hart_cuts_bias_means}, hard algorithms break down in this case, producing biased classification results. There are physical reasons to expect overlapping classes: First, the random inclination and orientation angles w.r.t. the line of sight induce a continuous transition of apparent axis ratios, apparent steepness of the radial light profiles and ratio of light coming from bulge and disk components. Second, observations of galaxies show that there are indeed transitional objects between different morphological types. For instance, there are transitional objects between early- and late-type galaxies in the ''green valley'' of the colour bimodality \citep[e.g.][]{Strateva2001,Baldry2004}, which is also reproduced in simulations \citep{Croton2006}. Hence, we have to draw the conclusion that hard algorithms are \textit{generically inappropriate} for analysing galaxy morphologies. This conclusion is backed up by practical experience, since even various specialists usually do not agree in hard visual classifications \citep[e.g.][]{Bamford2009}. In fact, the outcome of multi-person visual classifications becomes a probability distribution automatically.

Furthermore, our algorithm is a clustering algorithm, i.e. we do not specify the morphological classes a priori, but let the algorithm discover them. This approach is called ''unsupervised learning'' and it is the method of choice if we are uncertain about the type of objects we will find in a given data sample. In the context of clustering analysis classes are referred to as \textit{clusters}, and we adopt this terminology in this article.

\subsection{Probabilistic data-to-class assignment}

Let $O$ denote an object and $\vec x$ its parametrisation. Furthermore, let $c_k$ denote a single class out of $k=1,\ldots,K$ possible classes, then $\prob(c_k|\vec x)$ denotes the probability of class $c_k$ given the object $O$ represented by $\vec x$. This conditional probability $\prob(c_k|\vec x)$ is called the \textit{class posterior} and is computed using Bayes' theorem
\begin{equation}\label{eq:def_cluster_posterior}
\prob(c_k|\vec x) = \frac{\prob(c_k)\,\prob(\vec x|c_k)}{\prob(\vec x)} \;\textrm{.}
\end{equation}
The marginal probability $\prob(c_k)$ is called \textit{class prior} and $\prob(\vec x|c_k)$ is called \textit{class likelihood}. The denominator $\prob(\vec x)$ acts as a normalisation factor. The class prior and likelihood are obtained from a generative model (Sect. \ref{sect:bipartite_graph_model}). Prior and posterior satisfy the following obvious normalisation constraints
\begin{equation}\label{eq:normalisation_of_cluster_posteriors_priors}
\sum_{k=1}^K \prob(c_k) = 1 \quad\textrm{and}\quad \sum_{k=1}^K \prob(c_k|\vec x) = 1  \;\textrm{,}
\end{equation}
which ensure that each object is definitely assigned to some class. In the case of hard assignments, both posterior $\prob(c_k|\vec x)$ and likelihood $\prob(\vec x|c_k)$ are replaced by Kronecker symbols.

\section{Parametrising galaxy morphologies with shapelets \label{sect:shapelets}}
\subsection{Basis functions and expansion}

We parametrise galaxy morphologies in terms of shapelets \citep{Refregier2003}. Shapelets are a scaled version of two-dimensional Gauss-Hermite polynomials and form a set of complete basis functions that are orthonormal on the interval $[-\infty,\infty]$. A given galaxy image $I(\vec x)$ can be decomposed into a linear superposition of basis functions $B_{m,n}(\vec x/\beta)$, i.e.
\begin{equation}\label{eq:linear_superposition}
I(\vec x) = \sum_{m,n=0}^\infty c_{m,n}B_{m,n}(\vec x/\beta) \,\textrm{,}
\end{equation}
where the $c_{m,n}$ denote the expansion coefficients that contain the morphological information and $\beta>0$ denotes a scaling radius. In practice, the number of basis functions we can use is limited by pixel noise, such that the summation in Eq. (\ref{eq:linear_superposition}) stops at a certain maximum order $N_\textrm{max}<\infty$ which depends on the object's signal-to-noise ratio and resolution. This means Eq. (\ref{eq:linear_superposition}) is an approximation only,
\begin{equation}\label{eq:approximated_linear_superposition}
I(\vec x) \approx \sum_{m,n=0}^{N_\textrm{max}} c_{m,n}B_{m,n}(\vec x/\beta) \,\textrm{.}
\end{equation}
We use the C++ algorithm by \citet{Melchior2007} to estimate $N_\textrm{max}$, the scale radius and the linear coefficients, which was shown to be faster and more accurate than the IDL algorithm by \citet{Massey2005}.

\subsection{Problems with shapelet modelling}

\begin{figure*}
	\includegraphics[width=5.5cm]{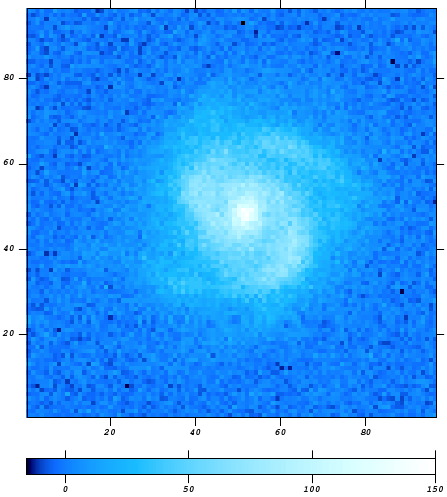}
	\includegraphics[width=5.5cm]{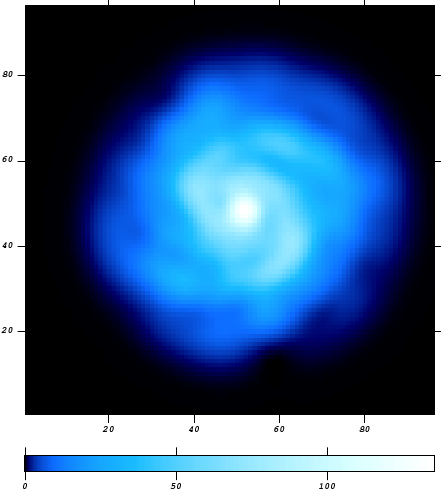}
	\includegraphics[width=5.5cm]{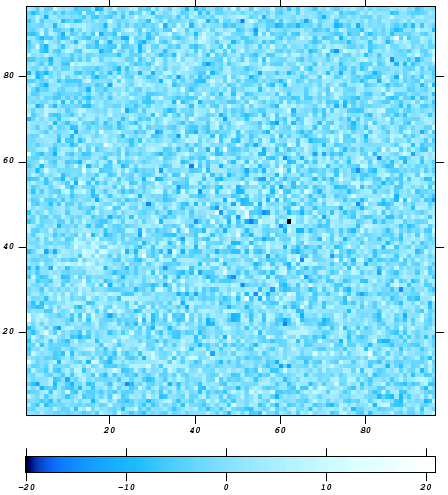}
\\*
	\includegraphics[width=5.5cm]{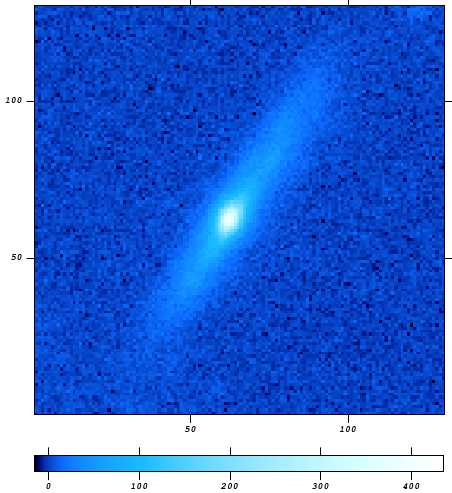}
	\includegraphics[width=5.5cm]{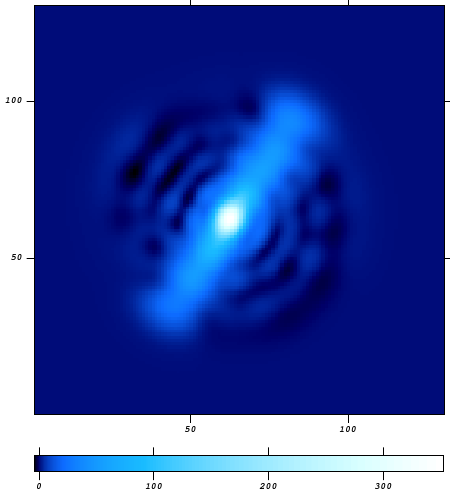}
	\includegraphics[width=5.5cm]{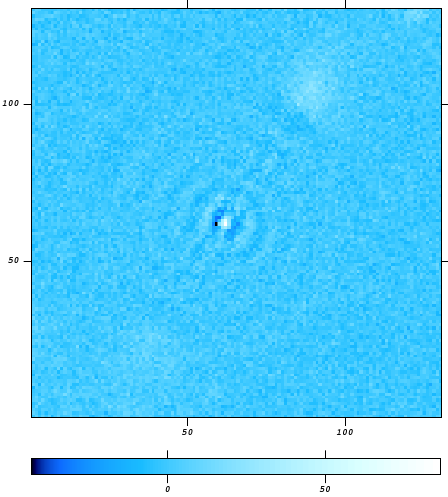}
\\*
	\includegraphics[width=5.5cm]{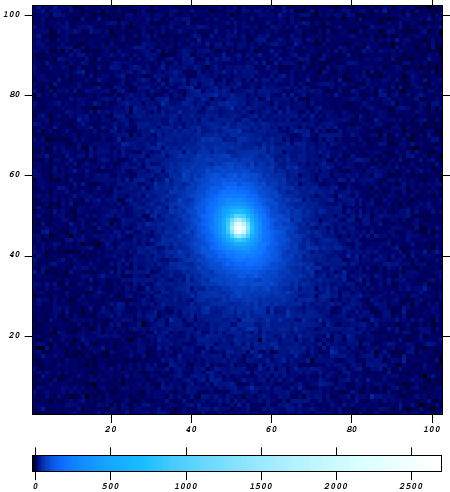}
	\includegraphics[width=5.5cm]{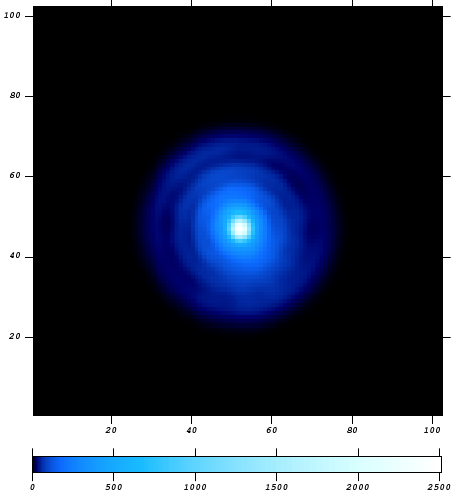}
	\includegraphics[width=5.5cm]{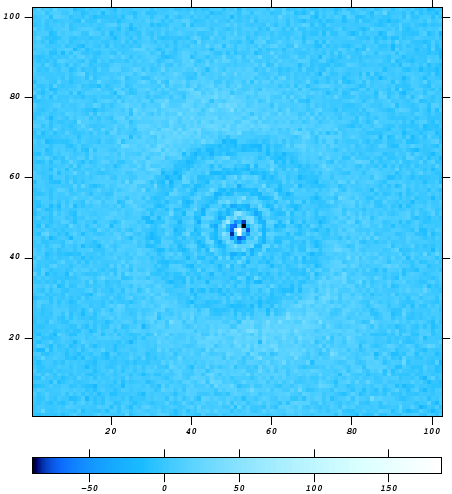}

\caption{Examples of shapelet models of three galaxies from SDSS ($g$ band).}
Shown are the original images (left column), the shapelet models (centre column) and the residuals (right column) of a face-on disk galaxy (top row), an edge-on disk galaxy (centre row) and an elliptical galaxy (bottom row). Note the different plot ranges of the residual maps. The shapelet decomposition used $N_\textrm{max}=16$, i.e. 153 basis functions.
\label{fig:example_shapelet_models_of_3_SDSS_galaxies}
\end{figure*}

It was shown by \citet{Melchior2009a} that the limitation of the number of basis functions in Eq. (\ref{eq:approximated_linear_superposition}) can lead to severe modelling failures and misestimations of galaxy shapes in case of objects with low signal-to-noise ratios. They identified two origins of these biases: First, the Gaussian profile of shapelets does not match the true profiles of galaxies, which are typically much steeper. Second, the shapelet basis functions are intrinsically spherical, i.e. they have problems in modelling highly eccentric objects. However, in this demonstration we consider only galaxies with high signal-to-noise ratios, where we can use many basis functions such that the impact of these biases is negligible. We demonstrate this in Fig. \ref{fig:example_shapelet_models_of_3_SDSS_galaxies}, where we show the shapelet reconstructions of a face-on disk, an edge-on disk and an elliptical galaxy drawn from the sample presented in Sect. \ref{sect:analysis_small_set}. The reconstruction of the face-on disk galaxy (top row) is excellent, leaving essentially uncorrelated noise in the residuals. However, the reconstructions of the edge-on disk galaxy (centre row) and the elliptical galaxy (bottom row) exhibit ring-like artefacts that originate from the steep light profiles of the elliptical and the edge-on disk along the minor axis. Such modelling failures appear systematically and do \textit{not} introduce additional scatter into the results, i.e. similar galaxies are affected in a similar way. However, since shapelet models do not capture steep and strongly elliptical galaxies very well, we are aware that our algorithm has less dicriminatory power for galaxies of this kind.

\subsection{Distances in shapelet space}

The coefficients form a vector space and we denote them as vectors $\vec x$. In first-order approximation, these coefficient vectors are independent of the size of the object, which was encoded by the scale radius $\beta$. Moreover, we can also make $\vec x$ invariant against the image flux, since Eq. (\ref{eq:linear_superposition}) implies that for a constant scalar $\alpha\neq 0$ the transformation $\vec x\rightarrow \alpha\vec x$ changes the image flux by this same factor of $\alpha$. Therefore, if we demand $\vec x\cdot\vec x=1$, then differing image fluxes will have no impact on the shapelet coefficients. This implies that morphologies are a \textit{direction} in shapelet coefficient space and the corresponding coefficient vectors lie on the surface of a hypersphere with unit radius. We can thus measure distances between morphologies on this surface via the angle spanned by their (normalised) coefficient vectors,
\begin{equation}\label{eq:def:spherical_distance}
d(\vec x_1,\vec x_2)=\sphericalangle\left(\vec x_1,\vec x_2\right) = \arccos\left(\vec x_1\cdot\vec x_2\right) \,\textrm{.}
\end{equation}
Employing the polar representation of shapelets \citep{Massey2005}, we can apply rotations and parity flips to shapelet models. We can estimate the object's orientation angle from the second moments of its light distribution \citep[e.g.][]{Melchior2007} and then use this estimate to align all models. This ensures invariance of the coefficients against random orientations. Additionally, we can break the degeneracy between left- and right-handed morphologies by applying parity flips such that the distance of two objects is minimised. These transformations in model space do not suffer from pixellation errors and increase the local density of similar objects in shapelet space.

\section{Soft Clustering Algorithm\label{sect:soft_clustering_algorithm}}

We now present the soft clustering algorithm of \citet{Yu2005}. Before we explain the details, we want to give a brief outline of the general method. The basic idea is to assign similarities to pairs of objects, so we first explain how to measure similarities of galaxy morphologies and what a similarity matrix is. These pairwise similarities are then interpreted by a probabilistic model, which provides our generative model. We also present the algorithm that fits the model to the similarity matrix.

\subsection{Estimating similarities \label{sect:optimal_similarity_measure}}

Instead of analysing the data in shapelet space, we compute a \textit{similarity matrix} by assigning similarities to any two data points. This approach is an alternative to working directly in the sparsely populated shapelet space or employing a method for dimensionality reduction. If we have $N$ data points $\vec x_n$, then this similarity matrix will be an $N\times N$ symmetric matrix. It is this similarity matrix to which we are going to apply the soft clustering analysis.

Based on the pairwise distances in shapelet coefficient space (Eq. (\ref{eq:def:spherical_distance})), we estimate pairwise similarities up to a constant factor as
\begin{equation}\label{eq:def:similarity_measure}
W_{mn} \propto 1 - \frac{(d(\vec x_m,\vec x_n)/d_\textrm{max})^\alpha}{s} \;\textrm{.}
\end{equation}
Here $d_\textrm{max}$ denotes the maximum distance between any two objects in the given data sample, while the exponent $\alpha>0$ and $s>1$ are free parameters that tune the similarity measure. We explain how to choose $\alpha$ and $s$ in Sect. \ref{sect:impact_non-optimal_simi}. This definition ensures that $0< W_{mn}\leq 1$ and that the maximum similarities are self-similarities for which $d(\vec x_m,\vec x_m)=0$. Note that this similarity measure is invariant under changes of size, flux, orientation, and parity of the galaxy morphology.

\subsection{Similarity matrices and weighted undirected graphs}

\begin{figure}\centering
  \begin{minipage}[t]{0.45\textwidth}
	\begin{minipage}[t]{1.0\textwidth}\centering
		\includegraphics[width=0.75\textwidth]{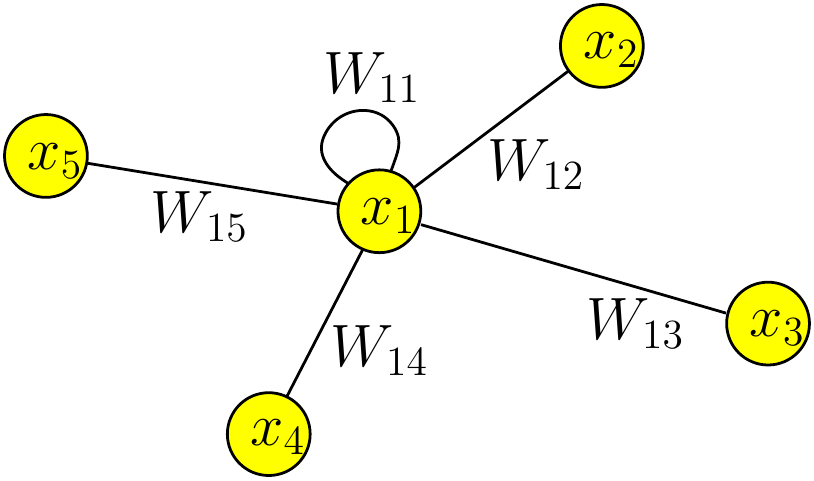}
	\end{minipage}
	\caption{Sketch of a weighted undirected graph.}
	\label{fig:similarity_matrix_as_graph}
The data nodes $x_n$ are connected by edges. For the sake of visibility, only edges connecting $x_1$ are shown. The edges are undirected and weighted by the similarity of the two connected nodes.
  \end{minipage}
\end{figure}

Square symmetric similarity matrices have a very intuitive interpretation: They represent a weighted undirected graph. Figure \ref{fig:similarity_matrix_as_graph} shows a sketch of such a graph. The data points $\vec x_n$ are represented symbolically as nodes $x_n$. The positions of these nodes are usually arbitrary, it is neither necessary nor helpful to arrange them according to the true locations of the data points in parameter space. Any two data nodes $x_m$ and $x_n$ are connected by an edge, which is assigned a weight $W_{mn}$. Obviously, all the weights $W_{mn}$ form an $N\times N$ matrix $ W$, and if this matrix is symmetric, i.e. $W_{mn}=W_{nm}$, the edges will have no preferred direction. In this case, the weighted graph is undirected. In graph theory the matrix of weights $ W$ is called \textit{adjacency matrix}, and we can interpret the similarity matrix as adjacency matrix of a weighted undirected graph.

Inspecting Fig. \ref{fig:similarity_matrix_as_graph}, we now introduce some important concepts. First, we note that there is also an edge connecting $x_1$ with itself. This edge is weighted by the ''self-similarity'' $W_{11}$. These self-similarities $W_{nn}$ are usually non-zero and have to be taken into account in order to satisfy normalisation constraints (cf. Eq. (\ref{eq:normalisation_of_W})). Second, we define the \textit{degree} $d_n$ of a data node $x_n$ as the sum of weights of all edges connected with $x_n$, i.e.
\begin{equation}
d_n = \sum_{m=1}^N W_{mn} \;\textrm{.}
\end{equation}
We can interpret the degree $d_n$ to measure the connectivity of data node $x_n$ in the graph. For instance, we can detect outlier objects by their low degree, since they are very dissimilar to all other objects. Third, we note that we can rescale all similarities by a constant scalar factor $C>0$ without changing the pairwise relations. Hence, we acquire the normalisation constraint
\begin{equation}\label{eq:normalisation_of_W}
\sum_{m,n=1}^N W_{mn} = \sum_{n=1}^N d_n = 1 \;\textrm{.}
\end{equation}
This constraint ensures the normalisation of the probabilistic model we are going to set up for our soft clustering analysis of the similarity matrix.

\subsection{Bipartite-graph model \label{sect:bipartite_graph_model}}

We need a probabilistic model of the similarity matrix $ W$ that can be interpreted in terms of a soft clustering analysis. Such a model was proposed by \citet{Yu2005}. As similarity matrices are closely related to graphs, this model is motivated from graph theory, too. The basic idea of this model is that the similarity of two data points $\vec x_m$ and $\vec x_n$ is induced by both objects being members of the same clusters. This is the basic hypothesis of any classification approach: Objects from the same class are more similar than objects from different classes.

\begin{figure}\centering
  \begin{minipage}[t]{0.45\textwidth}
	\begin{minipage}[t]{1.0\textwidth}\centering
		\includegraphics[width=0.5\textwidth]{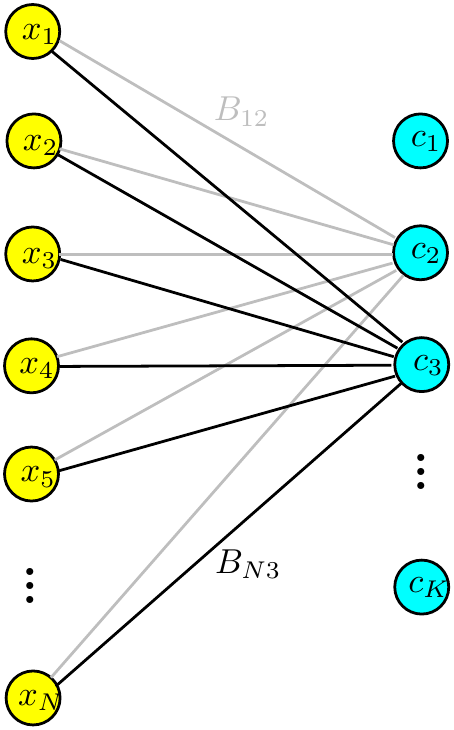}
	\end{minipage}
	\caption{Sketch of a bipartite graph.}
The bipartite graph contains two sets of nodes, $\mathcal X=\{x_1,\ldots,x_N\}$ and $\mathcal C=\{c_1,\ldots,c_K\}$. Edges connect nodes from different sets only and are weighted by an adjacency matrix $ B$. Not all edges are shown.
	\label{fig:bipartite_graph_sketch}
  \end{minipage}
\end{figure}

In detail, we model a weighted undirected graph (Fig. \ref{fig:similarity_matrix_as_graph}) and its similarity matrix by a \textit{bipartite graph} (Fig. \ref{fig:bipartite_graph_sketch}). A bipartite graph is a graph whose nodes can be divided into two disjoint sets $\mathcal X=\{x_1,\ldots,x_N\}$ of data nodes and $\mathcal C=\{c_1,\ldots,c_K\}$ of cluster nodes, such that the edges in the graph only connect nodes from different sets. Again, the edges are weighted and undirected, where the weights $B_{nk}$ form an $N\times K$ rectangular matrix $ B$, the bipartite-graph adjacency matrix. The bipartite-graph model for the similarity matrix then reads
\begin{equation}\label{eq:bipartite_graph_model}
\hat W_{mn} = \sum_{k=1}^K \frac{B_{nk}B_{mk}}{\lambda_k} \;\textrm{,}
\end{equation}
with the cluster priors $\lambda_k = \sum_{n=1}^N B_{nk}$. A detailed derivation is given in the following section. This model induces the pairwise similarities via two-hop transitions $\mathcal X\rightarrow\mathcal C\rightarrow\mathcal X$ \citep[cf.][]{Yu2005}. The numerator accounts for the strength of the connections of both data nodes to a certain cluster. The impact of the denominator is that the common membership to a cluster of small degree is considered more decisive. Obviously, the model defined by Eq. (\ref{eq:bipartite_graph_model}) is symmetric, as the similarity matrix itself. The normalisation constraint on $ W$ as given by Eq. (\ref{eq:normalisation_of_W}) translates via the bipartite-graph model to
\begin{equation}\label{eq:normalisation_B_and_lambda_k}
\sum_{k=1}^K \sum_{n=1}^N B_{nk} = \sum_{k=1}^K \lambda_k = 1 \;\textrm{.}
\end{equation}
These constraints need to be respected by the fit algorithm. Having fitted the bipartite-graph model to the given data similarity matrix, we compute the cluster posterior probabilities
\begin{equation}\label{eq:cluster_posteriors_bipartite-graph_model}
\prob(c_k|x_n) = \frac{\prob(x_n,c_k)}{\prob(x_n)} = \frac{B_{nk}}{\sum_{l=1}^K B_{nl}} \;\textrm{,}
\end{equation}
which are the desired soft data-to-cluster assignments. Obviously, $K$ cluster posteriors are assigned to each data node $x_n$ and the normalisation constraint $\sum_{k=1}^K\prob(c_k|x_n)=1$ is satisfied.

\subsection{Mathematical derivation \label{sect:deviation_bgm}}

Here we give a derivation of the bipartite-graph model of Eq. (\ref{eq:bipartite_graph_model}) that is more detailed than in \citet{Yu2005}. The ansatz is to identify the similarity $\hat W_{mn}$ with the joint probability
\begin{equation}\label{eq:W_as_joint_probability}
\hat W_{mn} = \prob(x_m,x_n) \;\textrm{.}
\end{equation}
This interprets $\hat W_{mn}$ as the probability to find $x_m$ and $x_n$ in the same cluster. Eq. (\ref{eq:normalisation_of_W}) ensures the normalisation $\sum_{m,n=1}^N \prob(x_m,x_n)=1$. As we do not know which particular cluster induces the similarity, we have to marginalise over all cluster nodes in Fig. \ref{fig:bipartite_graph_sketch},
\begin{equation}\label{eq:swtich_2_bipartite_graph_model}
\prob(x_m,x_n) = \sum_{k=1}^K \prob(x_m,x_n,c_k) \;\textrm{.}
\end{equation}
With this marginalisation we have switched from the weighted undirected graph to our bipartite-graph model. Applying Bayes' theorem yields
\begin{equation}
\prob(x_m,x_n) = \sum_{k=1}^K \prob(x_n|c_k)\,\prob(x_m,c_k) \;\textrm{,}
\end{equation}
where we have used
\begin{equation}\label{eq:assumption_in_derivation}
\prob(x_n|x_m,c_k) = \prob(x_n|c_k) \;\textrm{,}
\end{equation}
since $x_m$ and $x_n$ are not directly connected in the bipartite graph, i.e. they are statistically independent. This is the only assumption in this derivation and it implies that \textit{all} statistical dependence is induced by the clusters. Using Bayes' theorem once more yields
\begin{equation}
\prob(x_m,x_n) = \sum_{k=1}^K \frac{\prob(x_n,c_k)\,\prob(x_m,c_k)}{\prob(c_k)} \;\textrm{.}
\end{equation}
We identify the bipartite-graph adjacency matrix in analogy to  Eq. (\ref{eq:W_as_joint_probability}),
\begin{equation}\label{eq:def:bipartite_graph_adjacency_matrix}
B_{nk} = \prob(x_n,c_k) \;\textrm{,}
\end{equation}
with its marginalisation
\begin{equation}\label{eq:def_lambda_k}
\lambda_k = \prob(c_k) = \sum_{n=1}^N \prob(x_n,c_k) = \sum_{n=1}^N B_{nk} \;\textrm{.}
\end{equation}
The marginalised probabilities $\lambda_k$ are the cluster priors of the cluster nodes $c_k$ in the bipartite graph. Moreover, the $\lambda_k$ are the degrees of the nodes.

\subsection{Fitting the similarity matrix \label{sect:fitting_bipartite-graph_model}}

In order to fit the bipartite-graph model defined by Eq. (\ref{eq:bipartite_graph_model}) to a given similarity matrix, we perform some simplifications. First, we note that we can rewrite Eq. (\ref{eq:bipartite_graph_model}) using matrix notation,
\begin{equation}
\hat{ W} =  B\cdot\Lambda^{-1}\cdot B^T \;\textrm{,}
\end{equation}
where $ B$ is the $N\times K$ bipartite-graph adjacency matrix and $\Lambda=\textrm{diag}(\lambda_1,\ldots,\lambda_k)$ is the $K\times K$ diagonal matrix of cluster degrees. This notation enables us to employ fast and efficient algorithms from linear algebra. We change variables by
\begin{equation}
 B =  H\cdot\Lambda \;\textrm{,}
\end{equation}
where $ H$ is an $N\times K$ matrix. The elements of $ H$ can be interpreted as the cluster likelihoods, since $H_{nk}=\frac{B_{nk}}{\lambda_k}=\frac{\prob(x_n,c_k)}{\prob(c_k)}=\prob(x_n|c_k)$. Using these new variables $ H$ and $\Lambda$, the model $\hat{ W}$ of the data similarity matrix $ W$ is given by
\begin{equation}\label{eq:modified_bipartite-graph_model}
\hat{ W} =  H\cdot\Lambda\cdot H^T \;\textrm{,}
\end{equation}
where we have eliminated the matrix inversion and reduced the nonlinearity to some extent. The normalisation constraints of Eq. (\ref{eq:normalisation_B_and_lambda_k}) translate to $ H$ as
\begin{equation}\label{eq:normalisation_constraint_H}
\sum_{n=1}^N H_{nk} = \sum_{n=1}^N \prob(x_n|c_k) = 1 \qquad \forall\,k=1,\ldots,K \;\textrm{.}
\end{equation}
The normalisation constraints on $ H$ and $\Lambda$ are now decoupled, and we can treat both matrices as independent of each other. As $ H$ is an $N\times K$ matrix and $\Lambda$ a $K\times K$ diagonal matrix, we have $K(N+1)$ model parameters. In comparison to this number, we do have $\frac{1}{2}N(N+1)$ independent elements in the data similarity matrix due to its symmetry. Hence, a reasonable fit situation requires $\frac{1}{2}N\gg K$ in order to constrain all model parameters.

The data similarity matrix $ W$ is fitted by maximising the logarithmic likelihood function $\log\mathcal L$ of the bipartite-graph model. \citet{Yu2005} give a derivation of this function based on the theory of random walks on graphs. Their result is
\begin{equation}\label{eq:loglik_bipartite_graph_model}
\log\mathcal L(\Theta| W) = \sum_{m,n=1}^N W_{mn}\,\log\prob(x_m,x_n|\Theta) \;\textrm{,}
\end{equation}
where $\Theta=\{H_{11},\ldots,H_{NK},\lambda_1,\ldots,\lambda_K\}$ denotes the set of $K(N+1)$ model parameters and $\prob(x_m,x_n|\Theta)=\sum_{k=1}^K H_{mk}\lambda_k H_{nk}=\hat W_{mn}$ is the model. If we remember that $W_{mn}=\prob(x_m,x_n)$, then we see that $\log\mathcal L$ is the cross entropy of the true probability distribution $W_{mn}=\prob(x_m,x_n)$ and the modelled distribution $\hat W_{mn}=\prob(x_m,x_n|\Theta)$. Consequently, maximising $\log\mathcal L$ maximises the information our model contains about the data similarity matrix.

Directly maximising $\log\mathcal L$ is too hard, since the fit parameters are subject to the constraints given by Eqs. (\ref{eq:normalisation_B_and_lambda_k}) and (\ref{eq:normalisation_constraint_H}). We use an alternative approach that makes use of the expectation-maximisation (EM) algorithm, which is an iterative fit routine. Given an initial guess on the model parameters, the EM algorithm provides a set of algebraic update equations to compute an improved estimate of the optimal parameters that automatically respects the normalisation. These update equations are \citep{Bilmes1997,Yu2005}
\begin{equation}\label{eq:EM_update_lambda}
\lambda_k^\textrm{new} = \lambda_k \sum_{m,n=1}^N \frac{W_{mn}}{\left( H\cdot\Lambda\cdot H^T\right)_{mn}} H_{mk} H_{nk} \,\textrm{,}
\end{equation}
\begin{equation}\label{eq:EM_update_H}
H_{nk}^\textrm{new}\propto H_{nk}\lambda_k \sum_{m=1}^N  \frac{W_{mn}}{\left( H\cdot\Lambda\cdot H^T\right)_{mn}} H_{mk} \,\textrm{.}
\end{equation}
The $H_{nk}^\textrm{new}$ have to be normalised by hand, whereas the $\lambda_k^\textrm{new}$ are already normalised. Each iteration step updates all the model parameters, which has time complexity $\mathcal O(K\cdot N^2)$ for $K$ clusters and $N$ data nodes. We initialise all the cluster degrees to $\lambda_k^0=\frac{1}{K}$, whereby we trivially satisfy the normalisation condition and simultaneously ensure that no cluster is initialised as virtually absent. The $H_{nk}^0$ are initialised randomly and normalised ''by hand''.

Now, we want to briefly discuss the convergence properties of the EM algorithm. It has been shown \citep[e.g.][]{Redner1984} that the EM algorithm is guaranteed to converge to a \textit{local} maximum of $\log\mathcal L$ under mild conditions. Indeed, it was shown that the EM algorithm is monotonically converging, i.e. each iteration step is guaranteed to increase $\log\mathcal L$. Therefore, after each iteration step, we check how much $\log\mathcal L$ was increased compared to the previous step. If $\log\mathcal L$ changed by less than a factor of $10^{-9}$, we will consider the EM algorithm to have converged. This convergence criterion was chosen based on systematic tests like those discussed in Sect. \ref{sect:systematic_tests}. Finally, we note that the fit results are not unique, since the ordering of the clusters is purely random.

\subsection{Estimating the optimal number of clusters \label{sect:estimating_optimal_K}}

In this section we demonstrate how to estimate the optimal number of clusters for a given data set, which is a crucial part of any clustering analysis. It is essential to estimate the optimal number of clusters with due caution. This is a problem of assessing nonlinear models and there are no theoretically justified methods, there are only heuristic approaches. Common heuristics are the Bayesian information criterion
\begin{equation}\label{eq:def_BIC}
\textrm{BIC} = -2\,\log\mathcal L + p\,\log N
\end{equation}
and Akaike's information criterion
\begin{equation}
\textrm{AIC} = -2\,\log\mathcal L + 2p \,\textrm{,}
\end{equation}
where $p$ and $N$ denote the number of model parameters and the number of data points, respectively. As we have seen in Sect. \ref{sect:fitting_bipartite-graph_model}, the bipartite-graph model involves $K(N+1)$ model parameters. Consequently, BIC and AIC are not applicable, since $\log\mathcal L$ is not able to compensate for the large impact of the penalty terms. This inability of $\log\mathcal L$ is likely to originate from the sparse data population in the high-dimensional parameter space. Another tool of model assessment is cross-validation, but this is computationally infeasible in this case.

\begin{figure}
	\includegraphics[width=8cm]{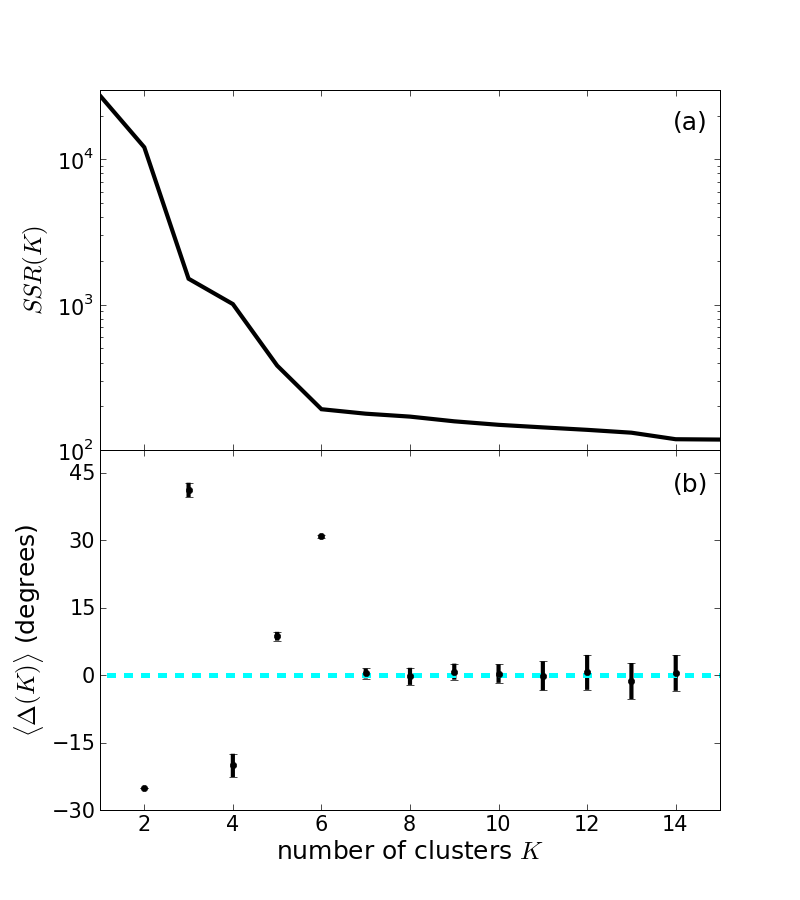}
\caption{Estimating the optimal number of clusters for the data sample shown in Fig. \ref{fig:demo_toy_example_data_set}.}
(a) $\SSR(K)$ as a function of the number $K$ of clusters. (b) Mean angular changes $\langle\Delta(K)\rangle$ averaged over ten fits.
\label{fig:demo_estimate_optimal_K}
\end{figure}

We now explain how to compare bipartite-graph models of different complexities heuristically, i.e. how to estimate the optimal number of clusters. This heuristic employs the sum of squared residuals
\begin{equation}\label{eq:definition_SSR}
\SSR(K) = \sum_{m=1}^N\sum_{n=1}^m \left( \frac{W_{mn} - \sum_{k=1}^K H_{mk}\lambda_k H_{nk} }{W_{mn}} \right)^2 \,\textrm{.}
\end{equation}
The definition puts equal emphasis on all elements. If we left out the denominator in Eq. (\ref{eq:definition_SSR}), the SSR would emphasise deviations of elements with large values, whereas elements with small values would be neglected. However, both large and small values of pairwise similarities are decisive. We estimate the optimal $K$ via the position of a \textit{kink} in the function $\SSR(K)$ (cf. Fig. \ref{fig:demo_estimate_optimal_K}). Such a kink arises if adding a further cluster does not lead to a significant improvement in the similarity-matrix reconstruction.

We demonstrate this procedure in Fig. \ref{fig:demo_estimate_optimal_K} by using the toy example of Figs. \ref{fig:demo_toy_example_data_set} and \ref{fig:demo_estimate_optimal_simi_measure}, which is composed of six nicely separable clusters. We fit bipartite-graph models to the similarity matrix shown in Fig. \ref{fig:demo_estimate_optimal_simi_measure}, with $K$ ranging from 1 to 15. The resulting SSR values are shown in panel (a) of Fig. \ref{fig:demo_estimate_optimal_K}. In fact, $\SSR(K)$ exhibits two prominent kinks at $K=3$ and $K=6$, rather than a single one. Obviously, for $K=3$, the clustering algorithm groups the four nearby clusters together, thus resulting in three clusters. For $K=6$, it is able to resolve this group of ''subclusters''.

We can construct a more quantitative measure by computing the angular change $\Delta(K)$ of $\log\SSR(K)$ at each $K$,
\begin{displaymath}
\Delta(K) = \arctan\left[\log\SSR(K-1) - \log\SSR(K)\right]
\end{displaymath}
\begin{equation}
 - \arctan\left[\log\SSR(K) - \log\SSR(K+1)\right] \;\textrm{.}
\end{equation}
As $K$ is an integer, $\log\SSR(K)$ is a polygonal chain and thus an angular change is well defined. A large positive angular change then indicates the presence of a kink in $\SSR(K)$.\footnote{It is not possible to compute the angular change for $K=1$, but this case is not a reasonable grouping anyway under the assumption that there are objects of different types in the given data sample.} However, we can even do better by fitting the similarity matrix several times for each $K$ and averaging the angular changes. The results of the fits differ slightly, since the model parameters are randomly initialised each time. These mean angular changes are shown in panel (b) of Fig. \ref{fig:demo_estimate_optimal_K}, averaged over 20 fits for each $K$. First, for large $K$ the mean angular changes are consistent with zero, i.e. in this domain increasing $K$ decreases $\SSR(K)$ but does not improve the fit systematically. Second, for $K=3$ and $K=6$ the mean angular changes deviate significantly from zero. For $K=2$ and $K=4$, the mean angular changes are negative, which corresponds to ''opposite'' kinks in the SSR spectrum and is due to $K=3$ being a very good grouping of the data.

For large $K$ these detections may be less definite due to the flattening of $\SSR(K)$. Therefore, we may systematically underestimate the optimal number of clusters. Moreover, this toy example also demonstrates that there may be more than a single advantageous grouping of the data and there may be disadvantageous groupings. If there are multiple detections of advantageous groupings, it may be difficult to judge which grouping is the best. In the worst case, we even may not find any signal of an advantageous grouping, which would either imply that our given sample is composed of objects of the same type or that the data does not contain enough information about the grouping. Unfortunately, this scheme of estimating the optimal number of clusters is extremely inefficient from a computational point of view. This is a severe disadvantage for very large data sets. Moreover, though this heuristic is working well, the significance of the mean angular changes is likely to be strongly influenced by the variance caused by the algorithm's initialisation.

\subsection{Comparison with previous work}

As the work of \citet{Kelly2004,Kelly2005} is very close to our own work, we want to discuss it in some detail and work out the differences. The authors applied a soft clustering analysis to the first data release of SDSS. In \citet{Kelly2004} they decomposed $r$-band images of 3,037 galaxies into shapelets, using the IDL shapelet code by \citet{Massey2005}. In \citet{Kelly2005} they extended this scheme to all five photometric bands $u,g,r,i,z$ of SDSS, thereby also taking into account colour information. Afterwards, they used a principal component analysis (PCA) to reduce the dimensionality of their parameter space. In \citet{Kelly2004} the reduction was from 91 to 9 dimensions and in \citet{Kelly2005} from 455 to 2 dimensions. Then they fitted a mixture-of-Gaussians model \citep{Bilmes1997} to the compressed data, where each Gaussian component represents a cluster. They were able to show that the resulting clusters exhibited a reasonable correlation to the traditional Hubble classes.

Reducing the parameter space with PCA and also using a mixture-of-Gaussians model are both problematic from our point of view. First, PCA relies on the assumption that those directions in parameter space that carry the desired information do also carry a large fraction of the total sample variance. This is neither guaranteed nor can it be tested for in practice. Second, galaxy morphologies are not expected to be normally distributed. Therefore, using a mixture-of-Gaussians model is likely to misestimate the data distribution. Nonetheless, the work by \citet{Kelly2004,Kelly2005} was a landmark, both concerning their usage of a probabilistic algorithm and conceptually, by applying a clustering analysis to the first data release of SDSS.

In contrast to \citet{Kelly2004,Kelly2005}, we do not reduce the dimensionality of the parameter space and then apply a clustering algorithm to the reduced data. We also do not try to model the data distribution in the parameter space, which would be virtually impossible due to its high dimensionality \citep[\textit{curse of dimensionality}, cf.][]{Bellman1961}. Rather, we use a similarity matrix, which has two major advantages: First, we do not rely on any compression technique such as PCA. Second, we cannot make any mistakes by choosing a potentially wrong model for the data distribution, since we model the similarity matrix. There are two sources of potential errors in our method: 
\begin{enumerate}
\item Estimation of pairwise similarities (Eq. (\ref{eq:def:similarity_measure})). This is hampered by our lack of knowledge about the metric in the morphological space and it is in some sense similar to mismodelling.
\item Modelling the similarity matrix by a bipartite-graph model. As the only assumption in the derivation of the bipartite-graph model is Eq. (\ref{eq:assumption_in_derivation}), this happens if and only if a significant part of the pairwise similarity is \textit{not} induced by the clusters, but rather by e.g. observational effects. However, any other classification method (automated or not) will have problems in this situation, too.
\end{enumerate}

\section{Systematic Tests\label{sect:systematic_tests}}

In this section we conduct systematic tests using artificial data samples that are specifically designed to investigate the impact of certain effects. First, we demonstrate that hard classification schemes cause problems with subsequent parameter estimation. Furthermore, we investigate the impact of non-optimal similarity measures, two-cluster separation, noise and cluster cardinalities on the clustering results.

\subsection{Overview}

We start by describing the artificial data sets that we are going to use. Furthermore, we describe the diagnostics by which we assess the performance of the clustering algorithm.

The data sets are always composed of two clusters, where the number of example objects drawn from each cluster may be different. The clusters are always designed as $p$-variate Gaussian distributions, i.e.
\begin{equation}
\prob(\vec x|\vec\mu,\Sigma) = \frac{\exp\left[ -\frac{1}{2}\left(\vec x-\vec\mu\right)^T\cdot\Sigma^{-1}\cdot\left(\vec x-\vec\mu\right) \right]}{\sqrt{\left(2\pi\right)^p\,\det\Sigma}} \;\textrm{,}
\end{equation}
where $\vec\mu$ and $\Sigma$ denote the mean vector and the covariance matrix, respectively.

Knowing the true analytic form of the underlying probability distributions, we are able to assess the probabilistic data-to-cluster assignments proposed by the clustering algorithm. For two clusters $A$ and $B$, the true data-to-cluster assignment of some data point $\vec x$ to cluster $k=A,B$ is given by the cluster posterior
\begin{equation}\label{eq:def_true_cluster_posteriors}
\prob(k|\vec x) = \frac{\prob(\vec x|\vec\mu_k,\Sigma_k)}{\prob(\vec x|\vec\mu_A,\Sigma_A) + \prob(\vec x|\vec\mu_B,\Sigma_B)} \;\textrm{.}
\end{equation}
The numerator $\prob(\vec x|\vec\mu_k,\Sigma_k)$ is the cluster likelihood. The cluster priors $\prob(A)=\prob(B)=\frac{1}{2}$ are flat and cancel out. For a given data set of $N$ objects, these true cluster posteriors are compared to the clustering results using the expectation values of the zero-one loss function
\begin{equation}
\langle\mathcal L_{01}\rangle = \frac{1}{N}\sum_{n=1}^N \left\{\begin{array}{lcl} 0 & \Leftrightarrow & \prob_\textrm{fit}(C_n|\vec x_n)\\
& & \;\;>\prob_\textrm{fit}(\neg C_n|\vec x_n) \\ 1 & \textrm{else} & \\ \end{array}\right.
\end{equation}
and of the squared-error loss function
\begin{equation}
\langle\mathcal L_\textrm{SE}\rangle = \frac{1}{N}\sum_{n=1}^N \left( \prob_\textrm{fit}(C_n|\vec x_n) - \prob_\textrm{true}(C_n|\vec x_n) \right)^2  \;\textrm{,}
\end{equation}
where $C_n$ denotes the correct cluster label of object $\vec x_n$ and $\neg C_n$ the false label. The zero-one loss function is the misclassification rate, whereas the squared-error loss function is sensitive to misestimations of the cluster posteriors that do not lead to misclassifications. As the two clusters are usually well separated in most of the following tests, the true maximum cluster posteriors are close to $100\%$. Therefore, misestimation means underestimation of the maximum posteriors, which is quantified by $\sqrt{\langle\mathcal L_\textrm{SE}\rangle}$.

\subsection{Impact of hard cuts on parameter estimation\label{sect:hart_cuts_bias_means}}

\begin{figure}
	\includegraphics[width=8cm]{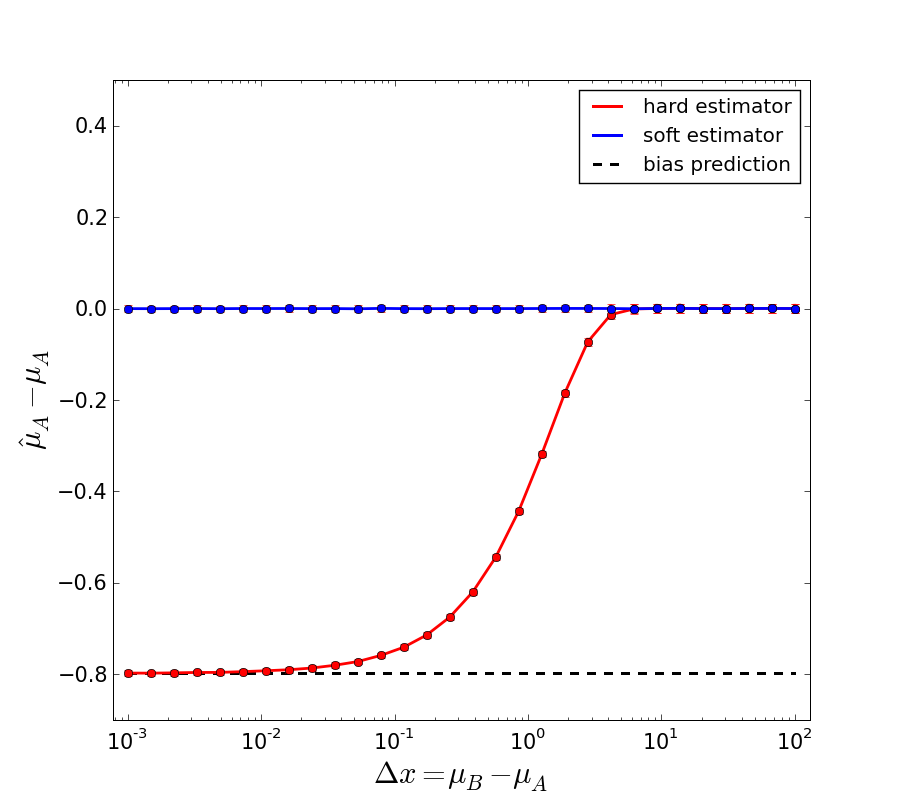}
\caption{Break-down of hard classifications in case of overlapping clusters.}
Deviation $\hat\mu_A-\mu_A$ of estimated and true means vs. two-cluster separation $\Delta x$ for class A for hard estimator (red line), soft estimator (blue line), and predicted bias of hard estimator for $\Delta x\rightarrow 0$ (dashed line). From 1,000 realisations of data samples we estimated errorbars, which are shown but too small to be visible.
\label{fig:hart_cuts_bias_means}
\end{figure}

In this first test, we want to demonstrate that hard cuts that are automatically introduced when using hard classification or clustering algorithms can lead to systematic misestimations of parameters, i.e. biases. This is a general comment in order to support our claim that hard data-to-class assignments are generically inappropriate for overlapping classes. We are not yet concerned with our soft-clustering algorithm. We use two one-dimensional Gaussians with means $\mu_A$ and $\mu_B$, variable two-cluster separation $\Delta x=\mu_A-\mu_B$, and constant variance $\sigma^2=1$. From each Gaussian cluster we then draw $N=$10,000 objects. From the resulting data sample we estimate the means $\hat\mu_k$ of the two Gaussians and compare with the true means $\mu_k$. The results are averaged over 1,000 realisations of data samples.

Figure \ref{fig:hart_cuts_bias_means} shows the deviations of the estimated from the true means when using a hard cut at $x=0$ (red line) and a weighted mean (blue line). A hard cut at $x=0$ that assigns all data points with $x<0$ to class A and those with $x>0$ to class B is the most reasonable hard classification in this simple example. Once the complete sample is divided into two subsamples for classes A and B, we estimate the usual arithmetic mean
\begin{equation}\label{eq:def:hard_estimator}
\hat\mu_k^\textrm{hard} = \frac{1}{N_k}\sum_{n=1}^{N_k} x_{k,n} \,\textrm{.}
\end{equation}
As Fig. \ref{fig:hart_cuts_bias_means} shows, this estimator is strongly biased in case of overlapping clusters ($\Delta x\rightarrow 0$). In the limit of $\Delta x = 0$, we can predict this bias analytically from the expectation value
\begin{equation}
\langle x\rangle_{A/B} = \mp 2\int_0^\infty dx\,x\,e^{-x^2/2} = \frac{\mp 2}{\sqrt{2\pi}} \approx \mp 0.7979 \,\textrm{,}
\end{equation}
where the integration is only over one half of the parameter space and the factor of 2 arises from both Gaussians contributing the same for $\Delta x = 0$. This bias is shown as dashed line in Fig. \ref{fig:hart_cuts_bias_means}, where for $\Delta x = 0$ also $\mu_{A/B}=\mp \Delta x/2=0$. If we employ the true posteriors defined by Eq. (\ref{eq:def_true_cluster_posteriors}) as weights and use
\begin{equation}\label{eq:def:soft_estimator}
\hat\mu_k^\textrm{soft} = \frac{\sum_{n=1}^N \prob(k|x_n)\,x_n}{\sum_{n=1}^N \prob(k|x_n)} \,\textrm{,}
\end{equation}
then we will get an unbiased estimate despite the overlap, as is evident from Fig. \ref{fig:hart_cuts_bias_means}. This comparison demonstrates the break-down of hard algorithms in case of overlapping clusters.

\subsection{Impact of non-optimal similarity measures \label{sect:impact_non-optimal_simi}}

\begin{figure}
\begin{center}
  \includegraphics[width=8cm]{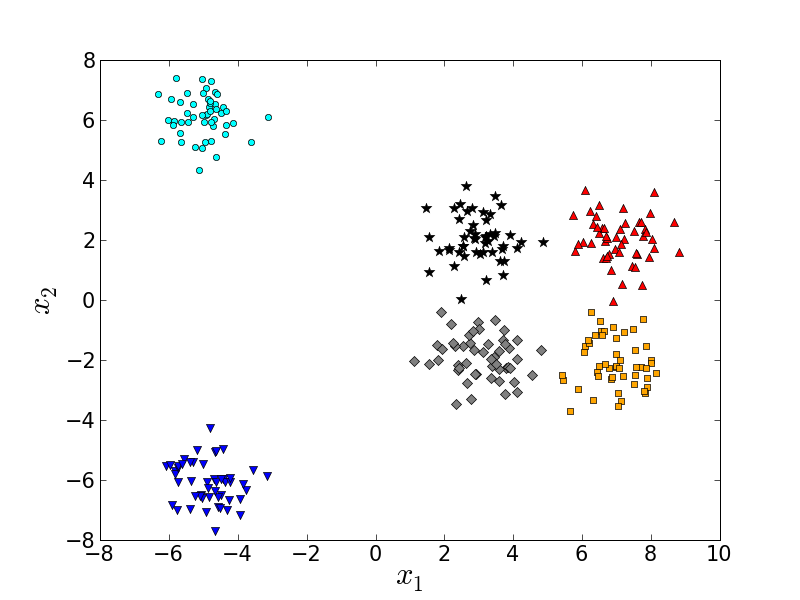}
  \includegraphics[width=8cm]{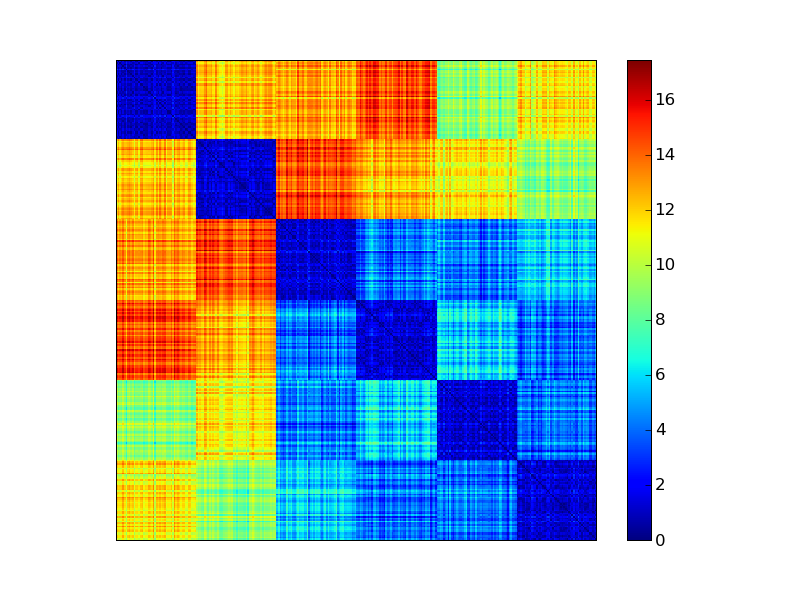}
\end{center}
\caption{Artificial data sample with six clusters (top) and the matrix of pairwise Euclidean distances (bottom).}
Each cluster has an underlying bivariate Gaussian distribution with covariance matrix $\Sigma=\diag(1,1)$. We sampled 50 data points from each cluster.
\label{fig:demo_toy_example_data_set}
\end{figure}

\begin{figure}
  \includegraphics[width=8cm]{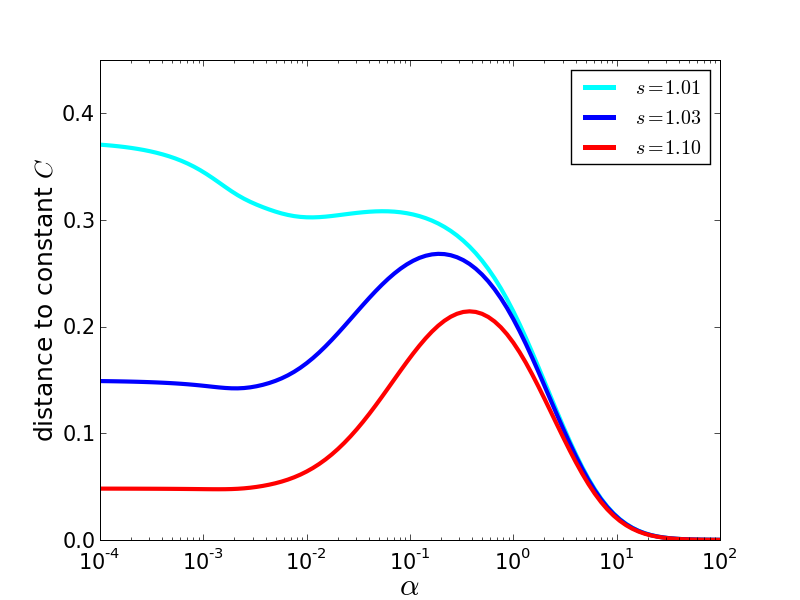}
  \includegraphics[width=8cm]{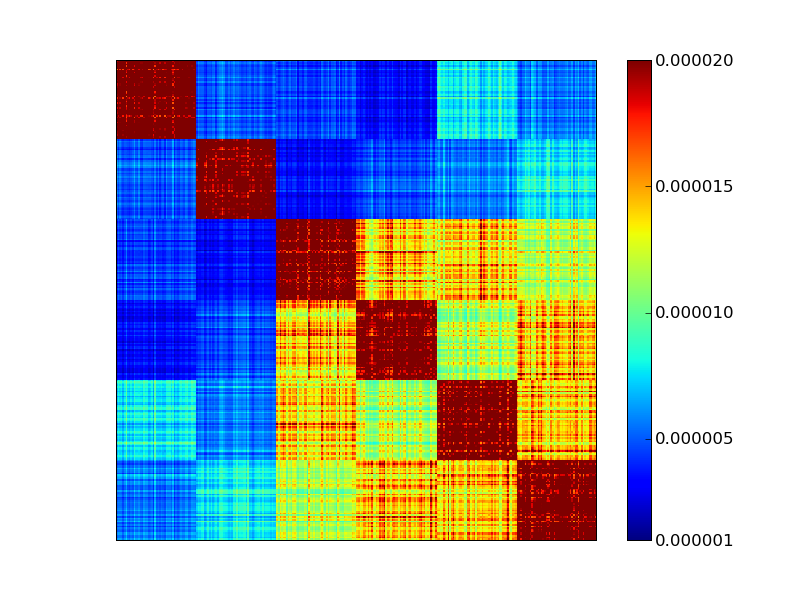}
\caption{Estimating the optimal similarity measure for the example data of Fig. \ref{fig:demo_toy_example_data_set}.}
Top panel: Modified Manhattan distance $C$ (Eq. (\ref{eq:def:Manhattan_dist_constant})) for $s=1.01$ (cyan line), $s=1.03$ (blue line) and $s=1.1$ (red line). For $\alpha\rightarrow 0$ the matrix becomes a step matrix, which is why the constant levels depend on the scale parameter. Bottom panel: The resulting similarity matrix.
\label{fig:demo_estimate_optimal_simi_measure}
\end{figure}

We now explain how to optimise the similarity measure defined in Eq. (\ref{eq:def:similarity_measure}) and what ''optimal'' means. Given the $N\times N$ symmetric matrix of pairwise distances $d(\vec x_m,\vec x_n)$, we can tune the similarity measure by adjusting the two parameters $\alpha$ and $s$. Tuning the similarity measure has to be done with care, since there are two undesired cases: First, for $\alpha\rightarrow\infty$, the resulting similarity matrix approaches a constant, i.e. $W_{mn}=\frac{1}{N^2}$ for all elements, since $d(\vec x_m,\vec x_n)\leq d_\textrm{max}$. This case prefers $K=1$ clusters, independent of any grouping in the data. Second, for $\alpha\rightarrow 0$, the similarity matrix approaches the step matrix defined by
\begin{equation}\label{eq:def:step_matrix}
S_{mn} \propto \left\{\begin{array}{lcl} 1 & \Leftrightarrow & m=n \\ 1-\frac{1}{s} & \Leftrightarrow & m\neq n \end{array}\right. \;\textrm{,}
\end{equation}
which is normalised such that $\sum_{m,n=1}^N S_{mn}=1$. This case prefers  $K=N$ clusters. The \textit{optimal} similarity measure should be as different as possible from these two worst cases. We choose $\alpha$ and $s$ such that the modified Manhattan distance to the constant matrix
\begin{equation}\label{eq:def:Manhattan_dist_constant}
C = \sum_{m=1}^N\sum_{n=1}^m \left|W_{mn}-\frac{1}{N^2}\right|
\end{equation}
is large. Figure \ref{fig:demo_estimate_optimal_simi_measure} demonstrates how to tune the similarity measure using the artificial data set from the toy example of Fig. \ref{fig:demo_toy_example_data_set}. The basis is the $N\times N$ symmetric distance matrix shown in the bottom panel of Fig. \ref{fig:demo_toy_example_data_set}. For three different values of $s$, the top panel shows $C$ as functions of $\alpha$. Obviously, $C(\alpha)$ exhibits a maximum and can thus be used to choose $\alpha$. For $s=1.1$ (red curve) the maximum is lowest and so is the distance to a constant matrix. $s=1.01$ exhibits the maximum deviation from a constant matrix, but this choice of $s$ downweights off-diagonal terms in $ W$ according to Eq. (\ref{eq:def:step_matrix}). Thus, we also prefer if $s$ is not too close to 1 and $s=1.03$ (blue curve) is the compromise of the three scale parameters shown in Fig. \ref{fig:demo_estimate_optimal_simi_measure}. Note that the choice of $s$ is not an optimisation but a heuristic. Although the artificial data set of Fig. \ref{fig:demo_toy_example_data_set} and its distance matrix are very special, we experienced that $C(\alpha)$ as shown in Fig. \ref{fig:demo_estimate_optimal_simi_measure} is representative for the general case.

The resulting similarity matrix is shown in the right panel of Fig. \ref{fig:demo_estimate_optimal_simi_measure} and exhibits a block-like structure, since we have ordered the data points in the set. This is just for the sake of visualisation and does  \textit{not} affect the clustering results. We clearly recognise six blocks along the diagonal, because the within-cluster similarities are always larger than the between-cluster similarities. Furthermore, we recognise a large block of four clusters in the bottom right corner that are quite similar to each other, whereas the remaining two clusters are more or less equally dissimilar to all other clusters. Consequently, the similarity matrix indeed represents all the features of the data set shown in Fig. \ref{fig:demo_toy_example_data_set}.

We now demonstrate first that the optimal similarity measure indeed captures the crucial information on the data and what happens if we do not use the optimal similarity measure. We use an artificial data set composed of two one-dimensional Gaussian clusters, both with unit variance and two-cluster separation of $\Delta x=3$. We sample 100 example objects from each cluster and compute the matrix of pairwise distances using the Euclidean distance measure. This data set and its distance matrix remain unchanged. For a constant parameter $s=2.25$, we vary the exponent $\alpha$ in the similarity measure defined by Eq. (\ref{eq:def:similarity_measure}). For each value of $\alpha$, we fit bipartite-graph models with $K=1$, 2 and 3 to the resulting similarity matrix, averaging the results over 15 fits each time.

\begin{figure*}
	\includegraphics[width=8cm]{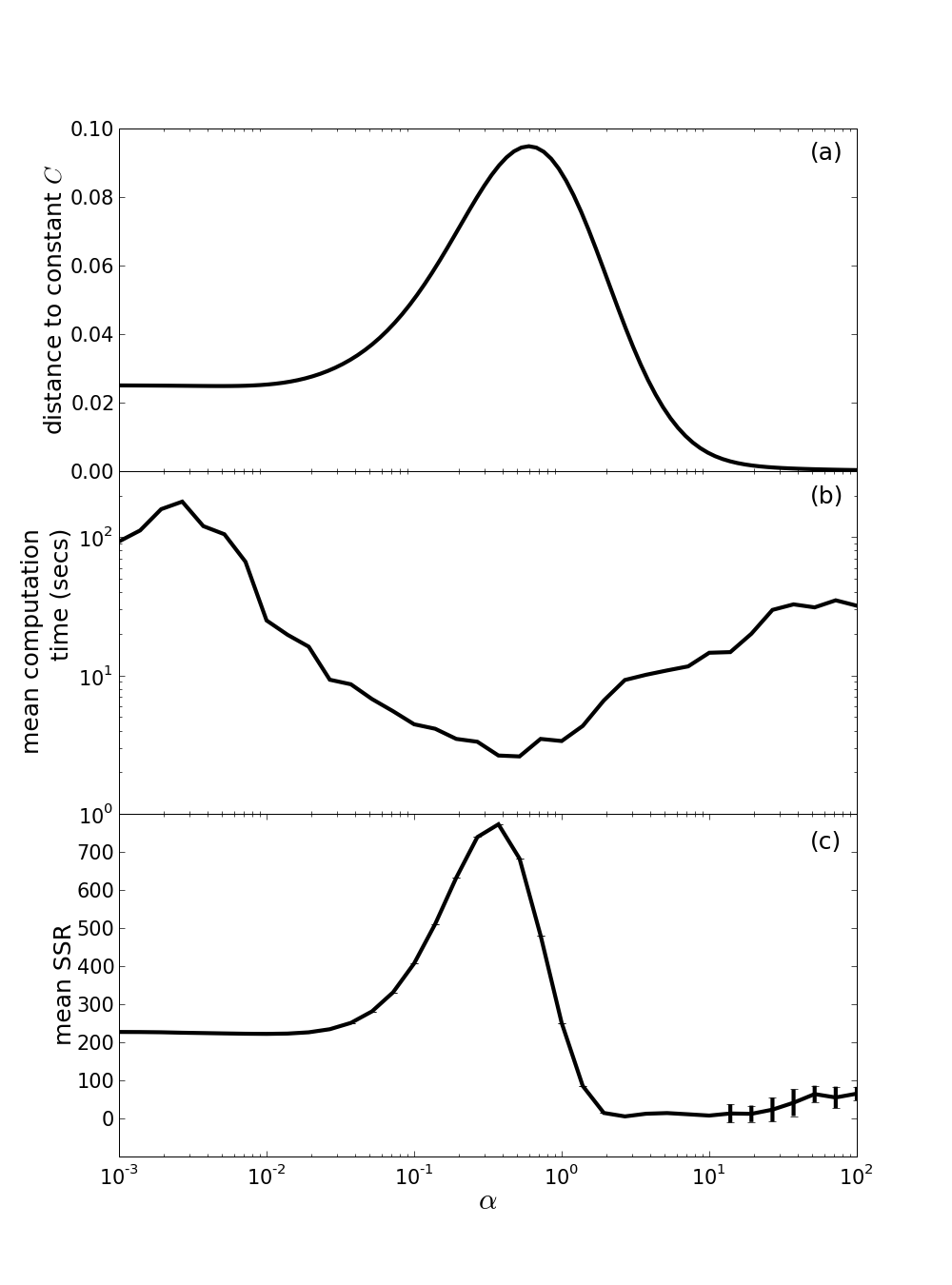}
	\includegraphics[width=8cm]{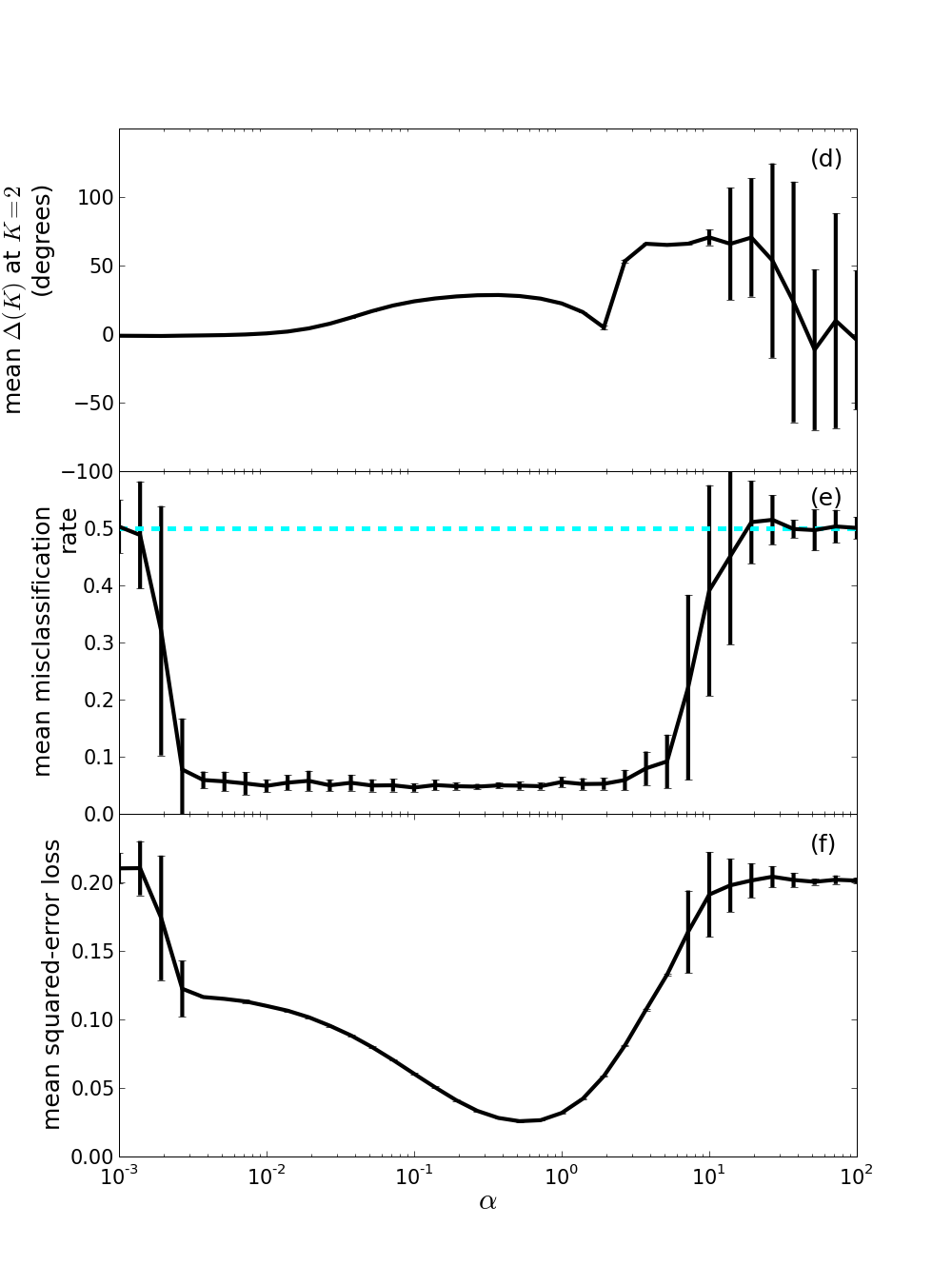}
\caption{Impact of non-optimal similarity measures on clustering results.}
(a) Modified Manhattan distance $C$ (Eq. (\ref{eq:def:Manhattan_dist_constant})). (b) Mean computation time per fit without errorbars. (c) Mean $\SSR(K)$ values of resulting fits for $K=2$. (d) Mean angular change of $\SSR(K)$ at $K=2$. (e) Mean misclassification rate (solid line) and 50\% misclassification rate (dashed line). (f) Mean squared-error loss of maximum cluster posteriors.
\label{fig:impact_non-optimal_simi_measure}
\end{figure*}

Results of this test are shown in Fig. \ref{fig:impact_non-optimal_simi_measure}. Panel (a) shows the modified Manhattan distance $C$ to a constant matrix. This curve is very similar to Fig. \ref{fig:demo_estimate_optimal_simi_measure}. There is a prominent peak at $\alpha\approx 0.6$, indicating the optimal similarity measure. If the similarity measure is very non-optimal, then the similarity matrix will be close to a constant or step  matrix, i.e. it poorly constrains the bipartite-graph model. In this case, we expect to observe overfitting effects, i.e. low residuals of the reconstruction and results with high variance. The computation times are longer, too, since the nonlinear model parameters can exhibit degeneracies thereby slowing down the convergence. Counterintuitively, we seek a large value of SSR in this test, since a similarity matrix which captures well the information content of the data is harder to fit. Indeed, the SSR values shown in panel (c) of Fig. \ref{fig:impact_non-optimal_simi_measure} are significantly lower for non-optimal $\alpha$'s and peak near the optimal $\alpha$. As expected, the mean computation times shown in panel (b) are minimal for the optimal similarity measure. Panel (d) shows how the evidence for two clusters evolves.\footnote{This is the reason why we need to fit bipartite-graph models using $K=1$, 2 and 3, in order to compute the angular change of $\SSR(K)$ at $K=2$.} Near the optimal $\alpha$, also the evidence for two clusters shows a local maximum. The misclassification rate shown in panel (e) is insensitive to $\alpha$ over a broad range, but approaches a rate of $50\%$ rather abruptly for extremely non-optimal similarity measures. The squared-error loss shown in panel (f) is more sensitive to non-optimalities. It exhibits a minimum for the optimal $\alpha$ and grows monotonically for non-optimal values.

The most important conclusion to draw from this test is that our method of choosing $\alpha$ and $s$ for the similarity measure defined in Sect. \ref{sect:optimal_similarity_measure} is indeed ''optimal'',  in the sense that it minimises both the misclassification rate and the squared-error loss. Additionally, we see that using the optimal similarity measure can also reduce computation times by orders of magnitude.

\subsection{Impact of two-cluster overlap \label{sect:impact_two-cluster_overlap}}

As we have to expect overlapping clusters in the context of galaxy morphologies, we now investigate the impact of the two-cluster overlap on the clustering results. The data sets used are always composed of 100 example objects drawn from two one-dimensional Gaussian clusters, both with unit variance. The two-cluster separation $\Delta x$ is varied from 1 to 1000. For each data set, we compute the matrix of pairwise Euclidean distances and then automatically compute the optimal similarity matrix by optimising $\alpha$ using a constant $s=2.25$ as described in Sect. \ref{sect:impact_non-optimal_simi}. To each similarity matrix we fit bipartite-graph models with $K=1$, 2 and 3 clusters. Furthermore, we fit a $K$-means algorithm with $K=1$, 2 and 3 to each data set in order to compare the results of both clustering algorithms. For each configuration, the results are averaged over 50 fits.

\begin{figure}
	\includegraphics[width=8cm]{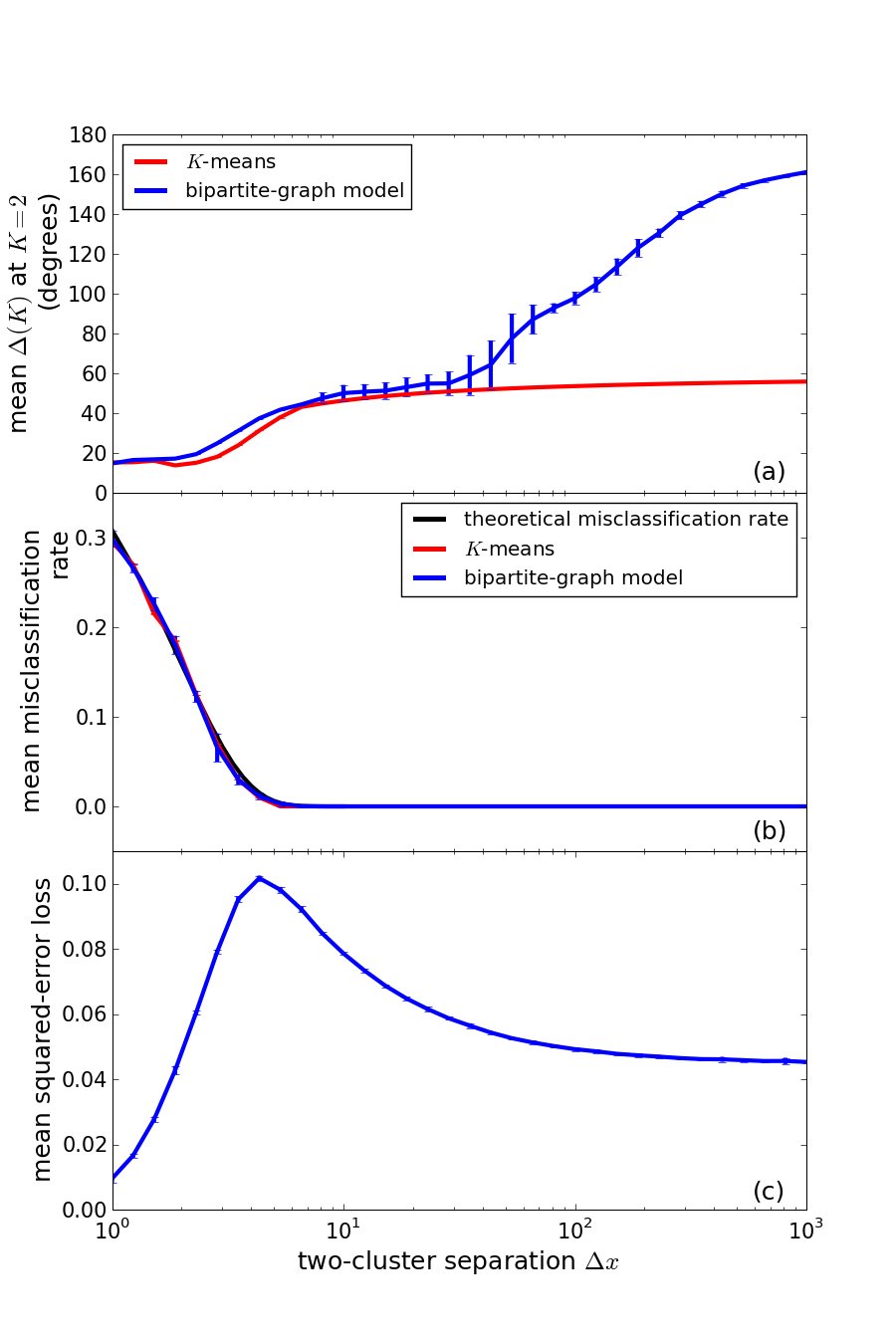}
	
\caption{Impact of two-cluster overlap on clustering results for $K$-means algorithm and bipartite-graph model.}
(a) Mean angular change of $\SSR(K)$ (bipartite-graph model) and within-cluster scatter ($K$-means) at $K=2$. (b) Mean misclassification rates of $K$-means and bipartite-graph model (see text) compared to theoretical prediction. All curves coincide. (c) Mean squared-error loss of bipartite-graph model.
\label{fig:impact_overlap}
\end{figure}

Results of this test are summarised in Fig. \ref{fig:impact_overlap}. Panel (a) shows the mean evidence for two clusters, based on the angular changes in $\SSR(K)$ for the bipartite-graph model and the within-cluster scatter for the $K$-means algorithm. For decreasing separation $\Delta x$, i.e. increasing overlap, the evidence for two clusters decreases for both algorithms, as is to be expected.\footnote{Note that these two curves cannot be compared directly. Their agreement for $\Delta x < 20$ is coincidence.} As panel (b) reveals, the misclassification rates for $K$-means and the bipartite-graph model are both in agreement with the theoretically expected misclassification rate expected in the ideal case (black curve). For two one-dimensional Gaussians with means $\pm\frac{\Delta x}{2}$, the theoretical misclassification rate is given by
\begin{equation}
\langle\mathcal L_{01}^\textrm{theo}\rangle = \int_{-\infty}^0 dx\,\prob\left(x\left|\mu=+\frac{\Delta x}{2}\right.,\sigma\right) \;\textrm{,}
\end{equation}
which measures the overlap of both Gaussians. In the limit $\Delta x=0$, this yields $\langle\mathcal L_{01}^\textrm{theo}\rangle = \frac{1}{2}$. The explanation for the excellent performance of both $K$-means and bipartite-graph model is, that in this case the clusters have equal cardinalities and are spherical. Nevertheless, the results of the $K$-means are biased due to the hard data-to-cluster assignment. Panel (c) of Fig. \ref{fig:impact_overlap} shows the mean squared-error loss of the bipartite-graph models.\footnote{We do not compare with $K$-means, since $K$-means is a hard algorithm and squared-error loss is no reasonable score function in this case.} First, the general trend is that the squared-error loss increases for decreasing two-cluster separation. This is due to the growing amount of overlap confusing the bipartite-graph model. Second, for $\Delta x\lesssim 4$, the squared-error loss decreases significantly. This effect can be explained as follows: For very small separations, the overlap is so strong that even the true cluster posteriors are both close to 50\%. Therefore, the fitted cluster posteriors scatter around 50\%, too, thereby reducing the squared error. Third, the squared error establishes a constant value of $\langle\mathcal L_\textrm{SE}\rangle\approx 0.045$ at large separations. In this case, the true maximum cluster posteriors are essentially $100\%$, so this corresponds to a systematic underestimation of the maximum posteriors of $\sqrt{\langle\mathcal L_\textrm{SE}\rangle}\approx 21\%$. Due to the large two-cluster separation, this \textit{bias} does not lead to misclassifications, as is evident from panel (b) in Fig. \ref{fig:impact_overlap}. This bias originates from the fact that any two objects have a finite distance and thus a non-vanishing similarity.

This test further demonstrates that the bipartite-graph model yields convincing results. This is most evident in the misclassification rate, which is in excellent agreement with the theoretical prediction of the best possible score.

\subsection{Impact of noise \label{sect:impact_noise}}

As observational data is subject to noise, we now investigate the response of the clustering results to noise on the similarity matrix. We simulate the noise by adding a second dimension $y$ to the data. The two clusters are bivariate Gaussian distributions, both with $\sigma_x^2=1$ and two-cluster separation of $\Delta x=10$ and $\Delta y=0$. We vary the size of the variance in $y$-direction ranging from $\sigma_y^2=0.1$ to 10000, thereby introducing noise that translates via the Euclidean distance to the similarity matrix. From each cluster 100 example objects are drawn and we fit bipartite-graph and $K$-means models using $K=1$, 2 and 3. The results are averaged over 50 fits for each value of $\sigma_y^2$.

\begin{figure}
	\includegraphics[width=8cm]{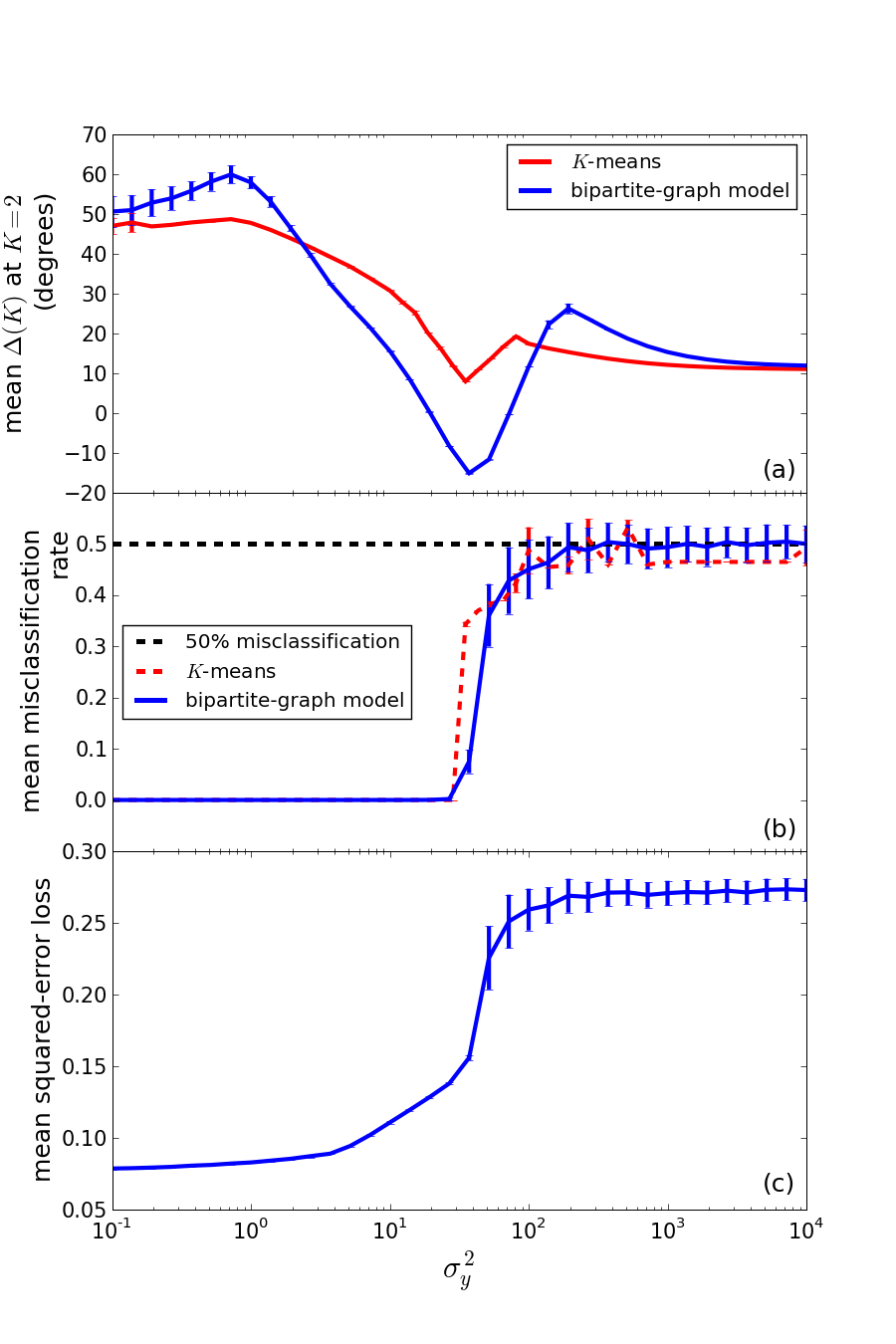}
\caption{Impact of noise variance $\sigma_y^2$ on clustering results for $K$-means algorithm and bipartite-graph model.}
(a) Mean angular change of $\SSR(K)$ (bipartite-graph model) and within-cluster scatter ($K$-means) at $K=2$. (b) Mean misclassification rate of $K$-means and bipartite-graph model. (c) Mean squared-error loss of bipartite-graph model.
\label{fig:impact_noise}
\end{figure}

Results of this test are shown in Fig. \ref{fig:impact_noise}. The evidence for two clusters (panel (a)) rapidly degrades for increasing variance for the bipartite-graph model as well as the $K$-means algorithm, as is to be expected. Inspecting the misclassification rate (panel (b)) reveals that both algorithms are insensitive to $\sigma_y^2$ until a critical variance is reached where both misclassification rates increase abruptly. For the $K$-means algorithm, this break down happens at $\sigma_y^2\approx 30$, whereas the bipartite-graph model breaks down at $\sigma_y^2\approx 40$, which amounts to $\frac{\Delta x}{\sigma_y}\approx 1.6$ in this setup. The evidence for two clusters (panel (a)) rises again for larger variances, although both algorithms have already broken down. This is a geometric effect: With increasing $\sigma_y^2$, the two clusters become more extended in $y$-direction, until it becomes favourable to split the data along $x=0$ rather than $y=0$. This also explains why the misclassification rate is 50\% in this regime. Consequently, the abrupt break-down originates from the setup of this test. Sampling more objects from each cluster might have prevented this effect, but would have increased the computational effort drastically. Moreover, this also demonstrates that isotropic distance measures are problematic. Using e.g. a diffusion distance \citep[e.g.][]{Richards2009} may solve this problem. The break down is less abrupt in the mean squared-error loss (panel (c)), since $\langle\mathcal L_{SE}\rangle$ is also sensitive to posterior misestimation that do not lead to misclassifications.

We conclude that the bipartite-graph model is fairly insensitive to noise of this kind over a broad range, until the setup of this test breaks down.

\subsection{Impact of cluster cardinalities\label{sect:focal_problem}}

Typically different types of galaxy morphologies have different abundancies in a given data sample. For instance, \citet{Bamford2009} observe different type fractions of early-type and spiral galaxies in the Galaxy Zoo project. Therefore, we now investigate how many objects of a certain kind are needed in order to detect them as a cluster. The concept of a number of objects being members of a certain cluster is poorly defined in the context of soft clustering. We generalise this concept by defining the \textit{cardinality} of a cluster $c_k$
\begin{equation}\label{eq:def:cluster_cardinality}
\textrm{card}(k) = \sum_{n=1}^N \prob(c_k|\vec x_n) \;\textrm{.}
\end{equation}
This definition satisfies $\sum_{k=1}^K \textrm{card}(k) = N$, since the cluster posteriors are normalised. In the case of hard clustering, Eq. (\ref{eq:def:cluster_cardinality}) is reduced to simple number counts, where the cluster posteriors become Kronecker symbols. We use two clusters, both one-dimensional Gaussians with unit variance and a fixed two-cluster separation of $\Delta x=10$. We then vary the number of objects drawn from each cluster such that the resulting data set always contains 200 objects. For each data set, we compute \textit{two different} similarity matrices: First, we compute the similarity matrix using the optimal $\alpha$ for a constant $s=2.0$, according to the recipe given in Sect. \ref{sect:impact_non-optimal_simi}. This similarity measure is adapted to every data set (\textit{adaptive similarity measure}). Second, we compute the similarity matrix using $\alpha=0.6$ and $s=2.0$, which is the optimal similarity measure for the data set composed to equal parts of objects from both clusters. This similarity measure is the same for all data sets (\textit{constant similarity measure}). To each of the two similarity matrices we fit a bipartite-graph model using $K=2$ and average the results over 50 fits.

\begin{figure}
	\includegraphics[width=8cm]{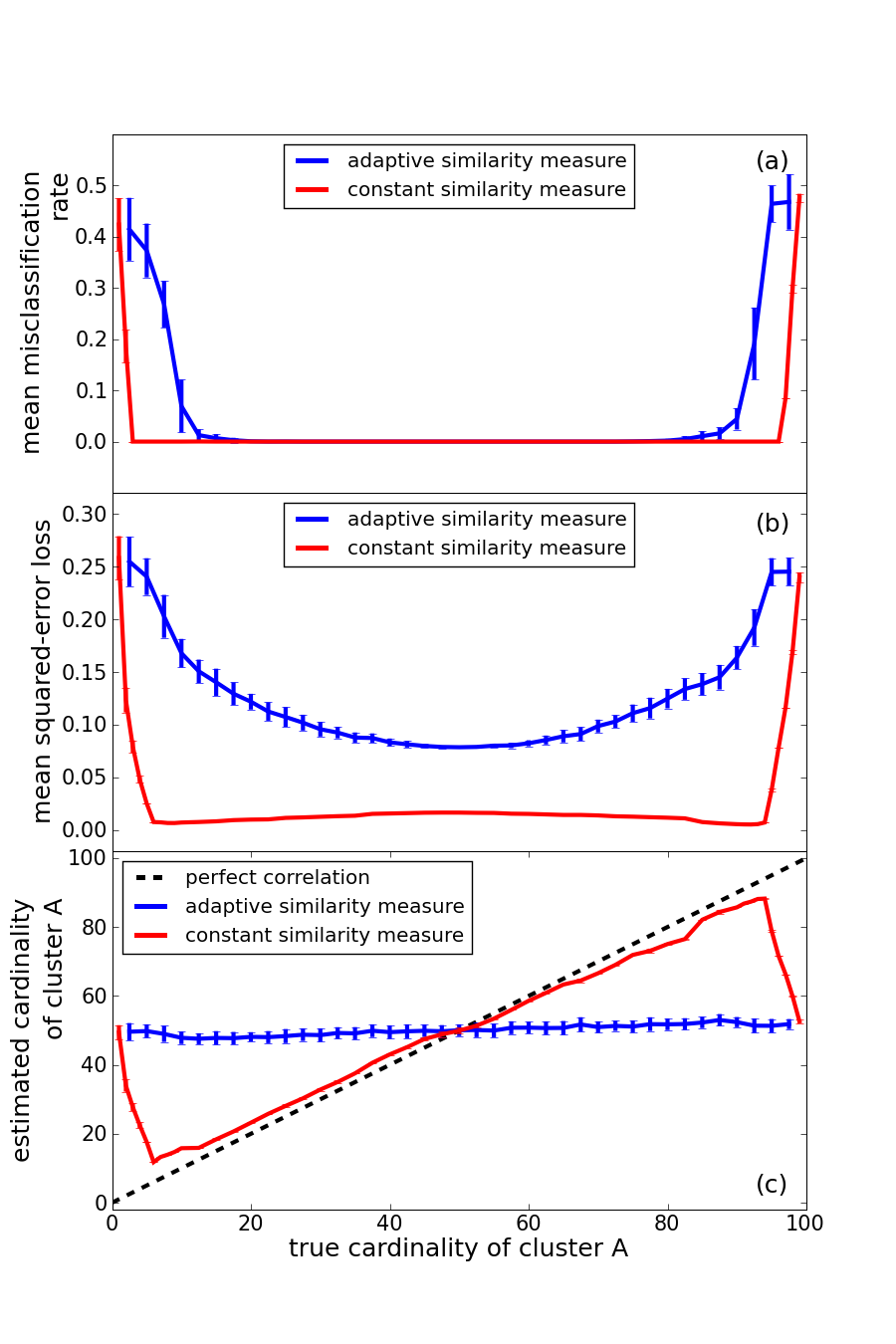}
\caption{Impact of cardinalities on clustering results.}
(a) Mean misclassification rate. (b) Mean squared-error loss. (c) Correlation of estimated and true cluster cardinality.
\label{fig:impact_cardinalities}
\end{figure}

The results are summarised in Fig. \ref{fig:impact_cardinalities}. Panel (a) shows the dependence of the misclassification rate on the cardinality of cluster A. For the adaptive similarity measure the bipartite-graph model will break down, if one cluster contributes less than $10\%$ to the data set. For the constant similarity measure it will break down, if one cluster contributes less than $3\%$. The same behaviour is evident from the squared-error loss in panel (b). This problem is caused by the larger group in the data set dominating the statistics of the modified Manhattan distance $C$ defined by Eq. (\ref{eq:def:Manhattan_dist_constant}). This is a failure of the similarity measure, \textit{not} of the bipartite-graph model. The constant similarity measure stays ''focussed'' on the difference between the two clusters and its break-down at $3\%$ signals the limit to which clusters are detectable with the bipartite-graph model.

Panel (c) in Fig. \ref{fig:impact_cardinalities} shows the correlation of the measured cluster cardinality to the true cluster cardinality. For the constant similarity measure, both quantities correlate well. In contrast to this, for the adaptive similarity measure the two quantities do not correlate at all. Again, the adaptive similarity measure is dominated by the larger group, i.e. the similarities between the large and the small group are systematically too high. This leads to a systematic underestimation of the maximum cluster posteriors (cf. panel (b)), since for a two-cluster separation of $\Delta x=10$ the true posteriors are essentially $100\%$ as shown by Fig. \ref{fig:impact_overlap}c. This also affects the cluster cardinalities defined by Eq. (\ref{eq:def:cluster_cardinality}). If the cluster overlap is stronger, then this bias is likely to lead to misclassifications, too.

We conclude that the optimal similarity measure defined in Sect. \ref{sect:impact_non-optimal_simi} fails to discover groups that contribute 10\% or less to the complete data sample. A different similarity measure may solve this problem, but the optimal similarity measure has the advantage of minimising the misclassification rate and the squared-error loss for the discovered groups.

\section{Worked Example with SDSS Galaxies\label{sect:worked_example_SDSS}}

In this section we present our worked example with SDSS galaxies. First, we describe the sample of galaxies we analyse. Before applying the bipartite-graph model to the whole sample, we apply it to a small subsample of visually classified galaxies to prove that it is working not only for simple simulated data but also for real galaxy morphologies. Again, we emphasise that this is just meant as a demonstration, so parametrisation and sample selection are idealised.

\subsection{The data sample by Fukugita et al. (2007) \label{sect:analysis_small_set}}

\citet{Fukugita2007} derived a catalogue of 2,253 bright galaxies with Petrosian magnitude in the $r$ band brighter than $r_P=16$ from the Third Data Release of the SDSS \citep{Abazajian2005}. We analyse only $g$-band imaging of this sample, which is sensitive to HII regions and spiral arm structures. We expect that objects that are bright in $r$ are also bright in the neighbouring $g$-band. Therefore, all these objects have a high signal-to-noise ratio, i.e. the shapelet decomposition can employ a maximum order sufficiently large to reduce possible modelling problems.

Apart from the $g$-band imaging data, we also retrieved further morphological information from the SDSS database, namely Petrosian radii $r_{50}$ and $r_{90}$ containing $50\%$ and $90\%$ of the Petrosian flux, ratios of isophotal semi major and semi minor axis, and the logarithmic likelihoods of best-fitting de Vaucouleurs and exponential-disk profiles. Given the Petrosian radii $r_{50}$ and $r_{90}$ containing $50\%$ and $90\%$ of the Petrosian flux, we define the concentration index in analogy to \citet{Conselice2003},
\begin{equation}\label{eq:concentration_index}
C = 5\,\log\left( \frac{r_{90}}{r_{50}} \right) \;\textrm{.}
\end{equation}
For compact objects, such as elliptical galaxies, this concentration index is large, whereas it is smaller for extended objects with slowly decreasing light profiles, such as disk galaxies.

We then reduce the data sample in three steps: First, we sort out peculiar objects, i.e. objects that are definitely not galaxies, blended objects and objects that were cut in the mosaic. All these objects have no viable galaxy morphologies. This was done by visual inspection of all objects. Second, we decompose all images into shapelets using the same maximum order $N_\textrm{max}=12$ (91 expansion coefficients) for all objects. The shapelet code performs several internal data processing steps, namely estimating the background noise and subtracting the potentially non-zero noise mean, image segmentation and masking of multiple objects, estimating the object centroid position \citep[cf.][]{Melchior2007}. Third, we sort out objects for which the shapelet reconstruction does not provide reasonable models. This is done by discarding all objects whose best fits have a reduced $\chi^2$ that is not in the interval $[0.9,2.0]$. The lower limit is chosen very close to unity, since shapelets have the tendency to creep into the background noise and overfit objects. Setting out from the 2,253 bright galaxies of \citet{Fukugita2007}, the data processing leaves us with 1,520 objects with acceptable $\chi^2$. We check that the morphological information contained in the original data set and the reduced data set does not differ systematically, by comparing the sample distributions of Petrosian radii, axis ratios, concentration indeces, and logarithmic likelihoods of best-fitting deVaucouleur and exponential-disk profiles. All objects are large compared to the point-spread function (PSF) of SDSS, such that a PSF deconvolution as described in \citet{Melchior2009} is not necessary. This means we analyse apparent instead of intrinsic morphologies, but both are approximately the same.

\subsection{Demonstration with three clusters \label{sect:testbed_FON_EON_ELL}}

In this section we apply the soft clustering algorithm by \citet{Yu2005} for the first time  to real galaxies. We use a small data set of 84 galaxies, which we visually classified as edge-on disk, face-on disk or ellipticals (28 objects per type). As these 84 galaxies were very large and very bright, we decomposed them anew using a maximum order of $N_\textrm{max}=16$, resulting in 153 shapelet coefficients per object. Figure \ref{fig:example_shapelet_models_of_3_SDSS_galaxies} shows one example object and its shapelet reconstruction for each type. This data set exhibits a strong grouping and we demonstrate that the bipartite-graph model indeed discovers the edge-on disks, face-on disks and ellipticals automatically, without any further assumptions.

\begin{figure}
      \includegraphics[width=8cm]{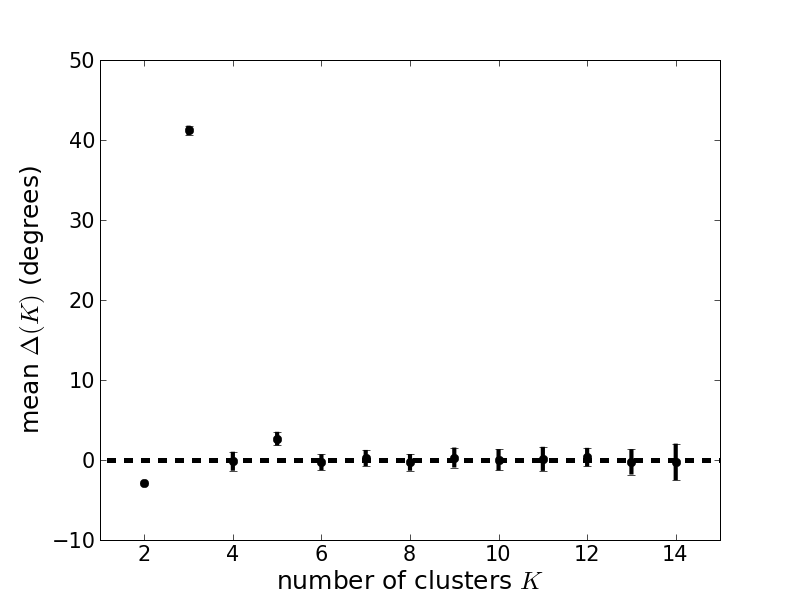}
\caption{Mean angular changes $\langle\Delta(K)\rangle$ of bipartite-graph model for data set composed of edge-on disks, face-on disks and ellipticals.}
\label{fig:3_class_testbed}
\end{figure}

The estimation of the number of clusters is shown in Fig. \ref{fig:3_class_testbed}. The mean angular changes in $\SSR(K)$ averaged over 20 fits indeed reveal only one significant  kink at $K=3$. The lowest value of SSR at $K=3$ is $\textrm{SSR}\approx 48$, which corresponds to an RMS residual (cf. Eq. (\ref{eq:definition_SSR})) of
\begin{equation}\label{eq:mean_residual_from_SSR}
\sqrt{\frac{\SSR}{ \frac{1}{2}N(N+1) }} \approx 11.6\% \;\textrm{.}
\end{equation}
The denominator $\frac{1}{2}N(N+1)$ is the number of independent elements in the symmetric similarity matrix.

\begin{figure}
      \includegraphics[width=8cm]{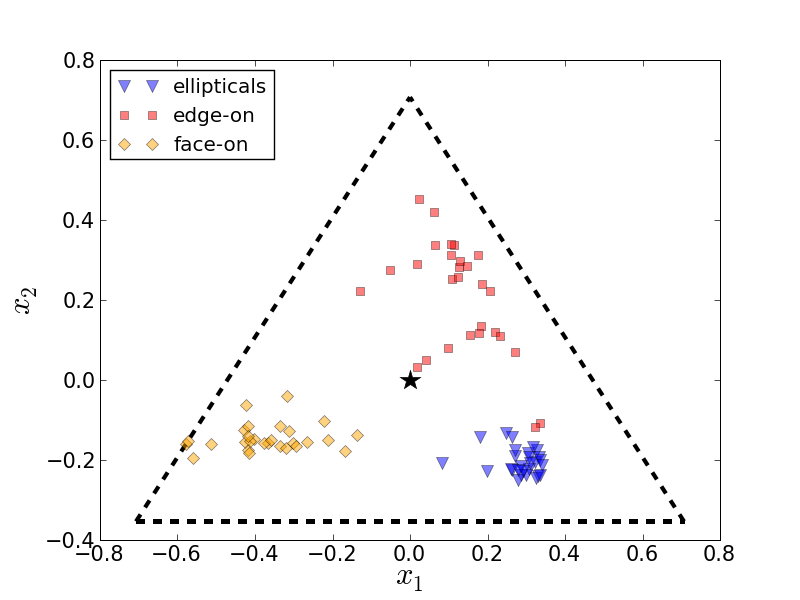}
\caption{Cluster posterior space of bipartite-graph model for edge-on disks, face-on disks and ellipticals.}
The triangle defines the subspace allowed by the normalisation constraint of the posteriors. The corners of the triangle mark the points of $100\%$ posterior probability. The * indicates the point where all three posteriors are equal. Colours encode a-priori classifications unknown to the algorithm.
\label{fig:3_class_testbed_posterior_planes}
\end{figure}

\begin{figure}
      \includegraphics[width=8cm]{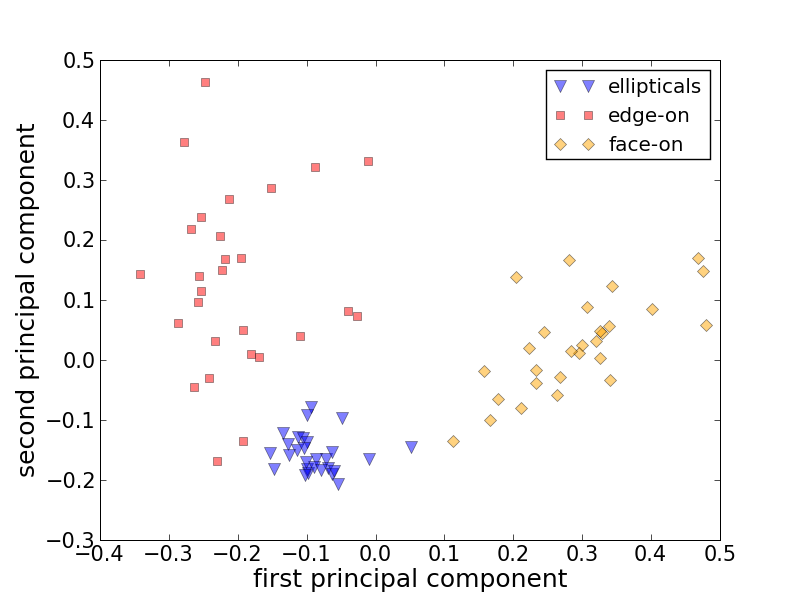}
\caption{Comparing Fig. \ref{fig:3_class_testbed_posterior_planes} with results of PCA for edge-on disks, face-on disks and ellipticals.}
Parameter space spanned by the first two principal components. The first principal component carries $\approx 45.2\%$ and the second $\approx 21.4\%$ of the total variance. Colours encode a-priori classifications unknown to the PCA algorithm.
\label{fig:3_class_PCA}
\end{figure}

We conclude from Fig. \ref{fig:3_class_testbed} that the bipartite-graph model indeed favours three clusters. However, we still have to prove that the similarity matrix contains sufficient information on the data and that the bipartite-graph model discovers the correct classes. For $K=3$, the cluster posteriors populate a two-dimensional plane because they are subject to a normalisation constraint. This plane is shown in Fig. \ref{fig:3_class_testbed_posterior_planes}. Indeed, the distribution of cluster posteriors exhibits an excellent grouping of ellipticals, edge-on disks and face-on disks. The three clusters are well separated, apart from two objects labelled as edge-on disks but assigned to the cluster of ellipticals. A second visual inspection of these two ''outliers'' revealed that we had initially misclassified them as edge-on disk. The excellent results are particularly impressive if we remember that we analysed 84 data points distributed in a 153-dimensional parameter space. Moreover, it is very encouraging that the soft clustering analysis did indeed recover the ellipticals, face-on and edge-on disks \textit{automatically}.

In order to get an impression of how good these results actually are, we compare the cluster posterior plane to results obtained from PCA. Therefore, we estimate the covariance matrix $\Sigma$ of the data sample in shapelet-coefficient space and diagonalise it. Only the first 83 eigenvalues of $\Sigma$ are non-zero, since the 84 data objects poorly constrain the $153\times 153$ covariance matrix. The first two principal components carry $66.6\%$ of the total sample variance and Fig. \ref{fig:3_class_PCA} displays the parameter space spanned by them. Obviously, PCA performs well in reducing the parameter space from 153 dimensions down to two, since the ellipticals, face-on and edge-on disks exhibit a good grouping.\footnote{PCA only reduces the parameter space, but does \textit{not} assign classes to objects.} However, the bipartite-graph model provides much more compact and well-separated groups. This is due to the degeneracies we have broken when we computed the minimal spherical distances as described in Sect. \ref{sect:shapelets}. In case of PCA, these degeneracies are unbroken and introduce additional scatter.

In both Figs. \ref{fig:3_class_testbed_posterior_planes} and \ref{fig:3_class_PCA} we notice that the group of ellipticals is significantly more compact than the groups of face-on and edge-on disks. This is caused by three effects: First, as discussed in Sect. \ref{sect:shapelets}, our parametrisation of elliptical galaxies is problematic, thereby introducing common artefacts for all objects of this type. These common features are then picked up by the soft-clustering algorithm. Ironically, the problems of the parametrisation help to discriminate the types in this case. Second, we described in Sect. \ref{sect:shapelets} how to make our morphological distance measure invariant against various random quantities, namely image size, image flux, orientation angle and handedness. However, the distance measure and thereby the similarity measure are \textit{not} invariant against the inclination angle w.r.t. the line of sight, which introduces additional scatter into the clustering results. We expect that the impact of this random effect is smaller for ellipticals than for disk galaxies. Third, disk galaxies usually exhibit complex substructures (e.g. spiral arms or star-forming regions), whereas elliptical galaxies do not. Consequently, the intrinsic morphological scatter of disk galaxies is larger than for ellipticals.

\subsection{Analysing the data set of Fukugita et al. (2007) \label{sect:analysing_Fuku}}

\begin{figure}
      \includegraphics[width=8cm]{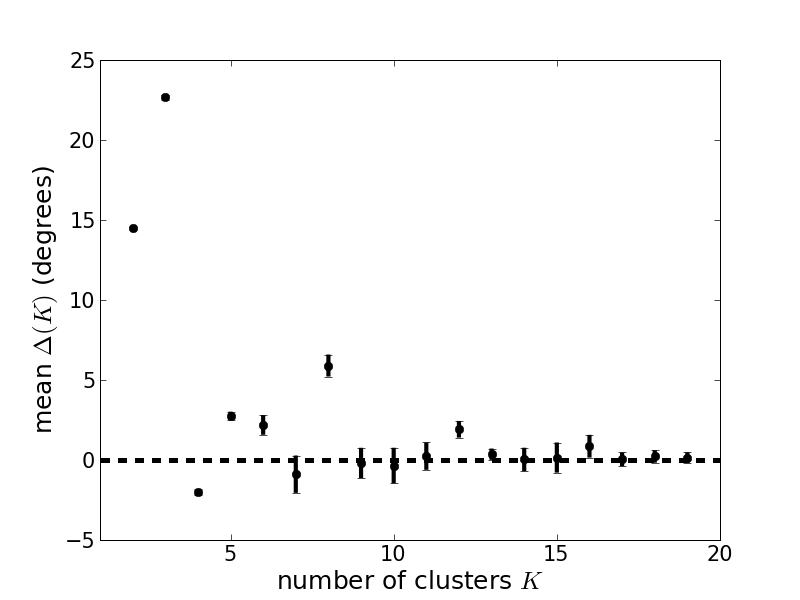}
\caption{Estimating the number of clusters in the data set of \citet{Fukugita2007}.}
Mean angular changes $\langle\Delta(K)\rangle$ are averaged over 15 Fits.
\label{fig:results_Fuku}
\end{figure}

We now present the soft-clustering results from analysing all 1,520 bright galaxies from the reduced data set of \citet{Fukugita2007}. We have chosen the similarity measure with $s=1.02$ and corresponding optimal $\alpha\approx 0.12$, according to Sect. \ref{sect:impact_non-optimal_simi}. The shapes of the curves of the modified Manhattan distances $C(\alpha)$ are of the same generic form as before. Fit results of the similarity matrix for $K$ ranging from 1 to 20 are shown in Table \ref{table:fit_results_small_sample} and Fig. \ref{fig:results_Fuku}. There are significant deviations of the mean angular changes from zero for $K=3$ and $K=8$. The signal at $K=2$ is ignored, since the SSR value is very high (cf. Table \ref{table:fit_results_small_sample}).

\begin{table}
\begin{tabular}{ccc}
\hline\hline
$K$ & minimal SSR & mean angular changes (degrees) \\
\hline
1  & $39,220$ & -- \\
2  & $12,313$ & $14.489\pm 0.047$ \\
3  & $6,146$ & $22.67\pm 0.14$ \\
4  & $4,965$ & $-2.01\pm 0.19$ \\
5  & $3,868$ & $2.76\pm 0.27$ \\
6  & $3,155$ & $2.19\pm 0.61$ \\
7  & $2,676$ & $-0.89\pm 1.17$ \\
8  & $2,254$ & $5.91\pm 0.69$ \\
9  & $2,093$ & $-0.18\pm 0.95$ \\
10 & $1,931$ & $-0.35\pm 1.11$ \\
11 & $1,790$ & $0.24\pm 0.86$ \\
12 & $1,661$ & $1.92\pm 0.52$ \\
13 & $1,593$ & $0.36\pm 0.35$ \\
14 & $1,532$ & $0.03\pm 0.73$ \\
15 & $1,476$ & $0.15\pm 0.94$ \\
16 & $1,430$ & $0.86\pm 0.71$ \\
17 & $1,405$ & $0.04\pm 0.43$ \\
18 & $1,383$ & $0.22\pm 0.39$ \\
19 & $1,365$ & $0.14\pm 0.34$ \\
20 & $1,348$ & -- \\
\hline
\end{tabular}
\caption{Fitting the similarity matrix of 1,520 objects.\newline We present the minimal SSR value out of 15 fits and the mean angular change averaged over 15 fits.}
\label{table:fit_results_small_sample}
\end{table}

\begin{figure}
      \includegraphics[width=1.6cm]{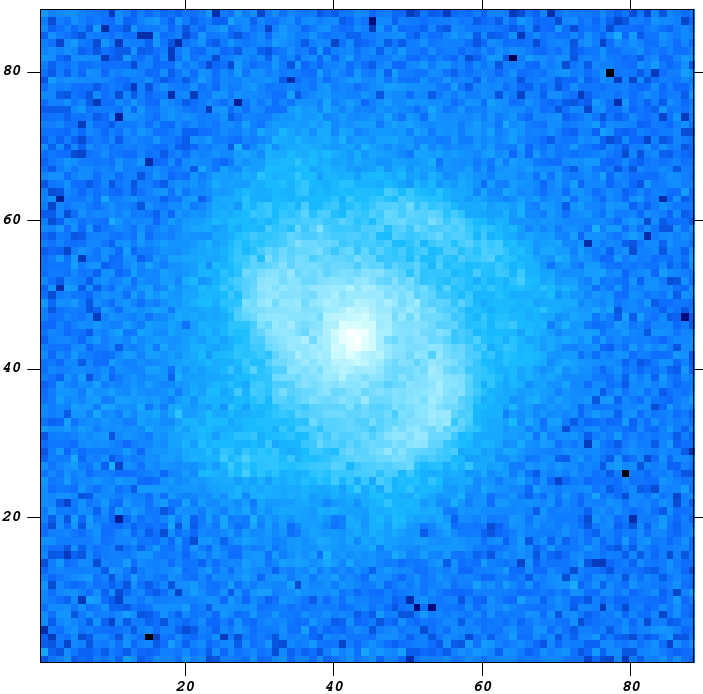}
      \includegraphics[width=1.6cm]{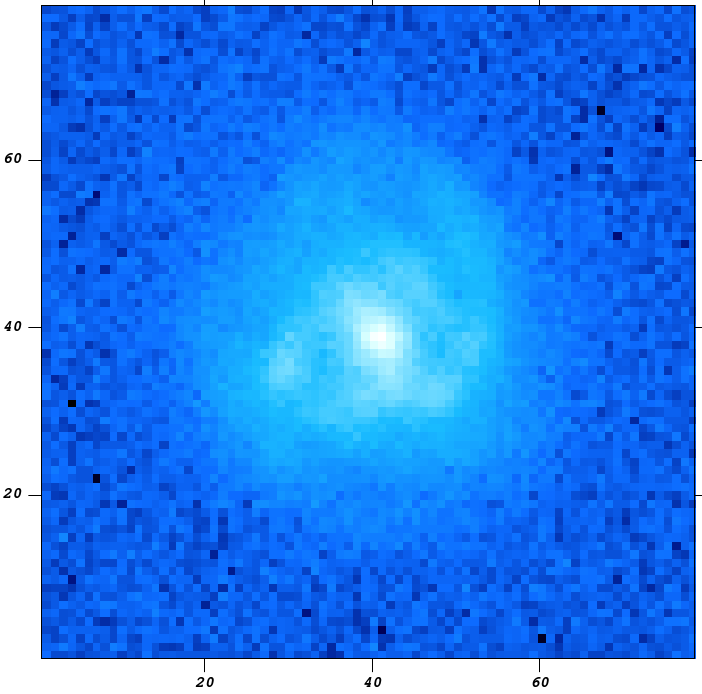}
      \includegraphics[width=1.6cm]{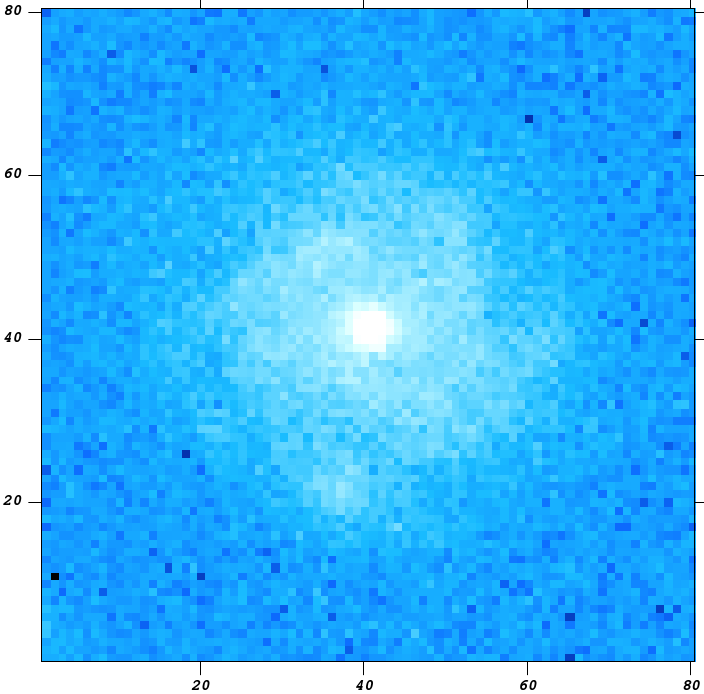}
      \includegraphics[width=1.6cm]{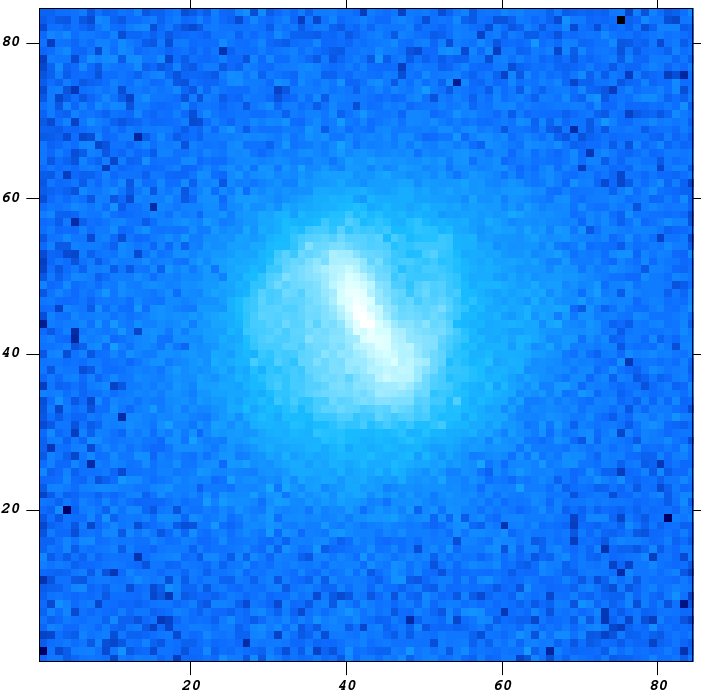}
      \includegraphics[width=1.6cm]{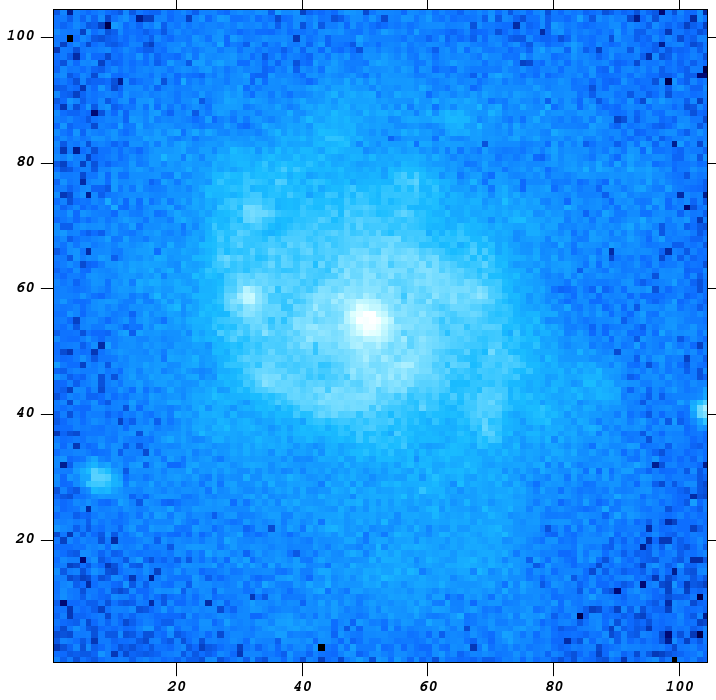}
${}$\\*
      \includegraphics[width=1.6cm]{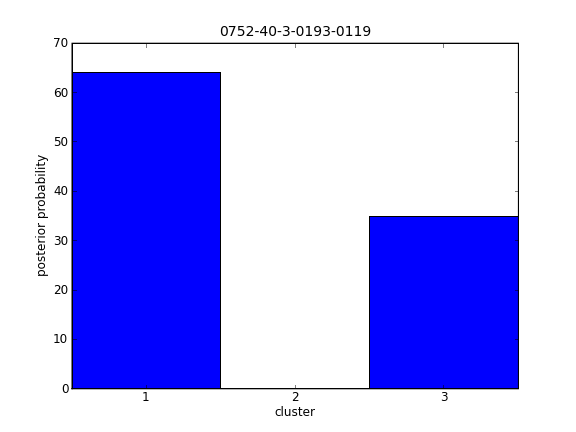}
      \includegraphics[width=1.6cm]{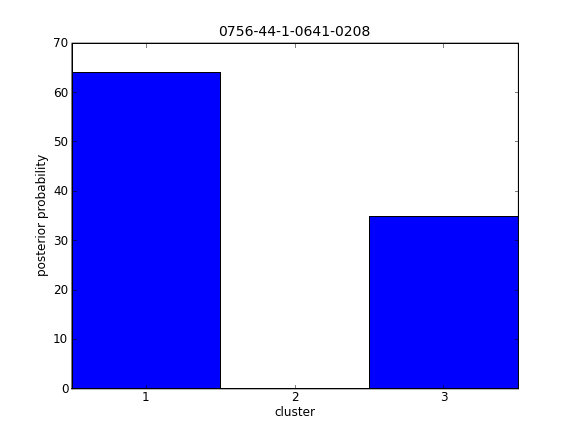}
      \includegraphics[width=1.6cm]{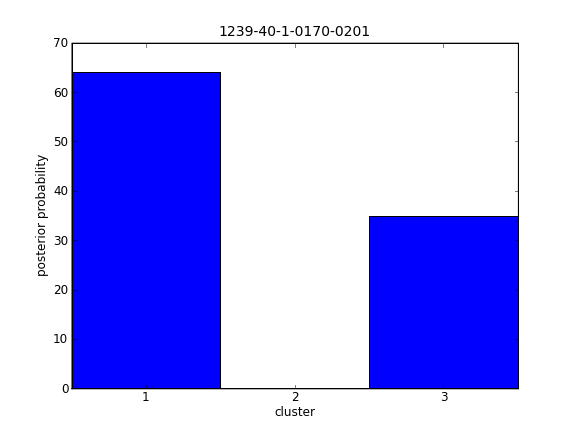}
      \includegraphics[width=1.6cm]{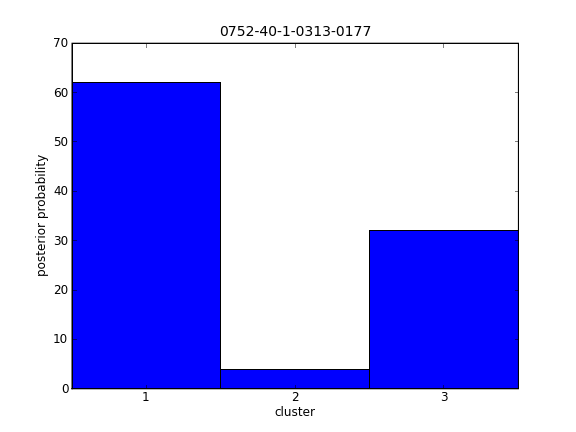}
      \includegraphics[width=1.6cm]{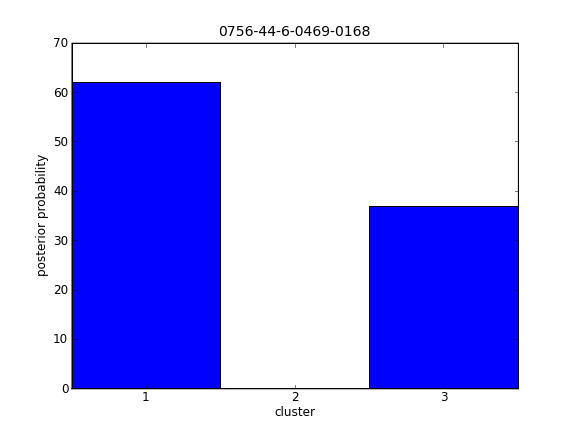}
${}$\\*
      \includegraphics[width=1.6cm]{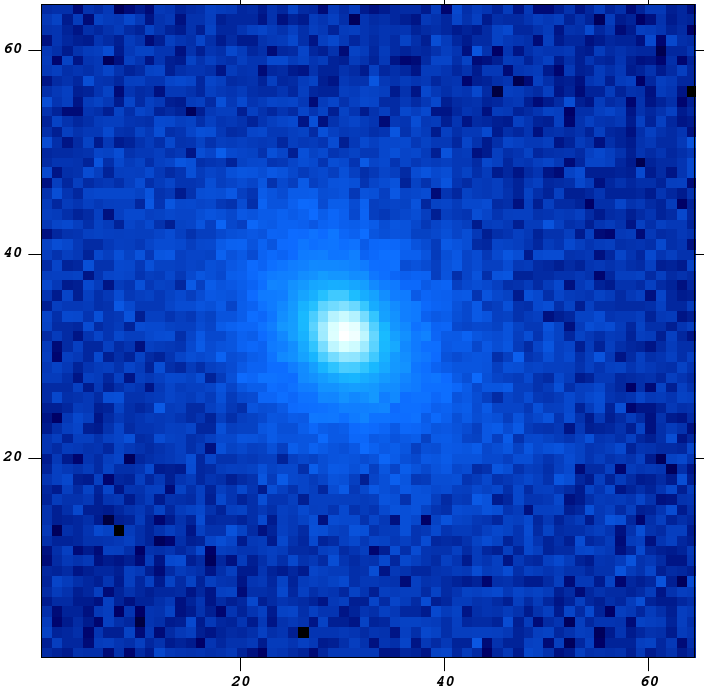}
      \includegraphics[width=1.6cm]{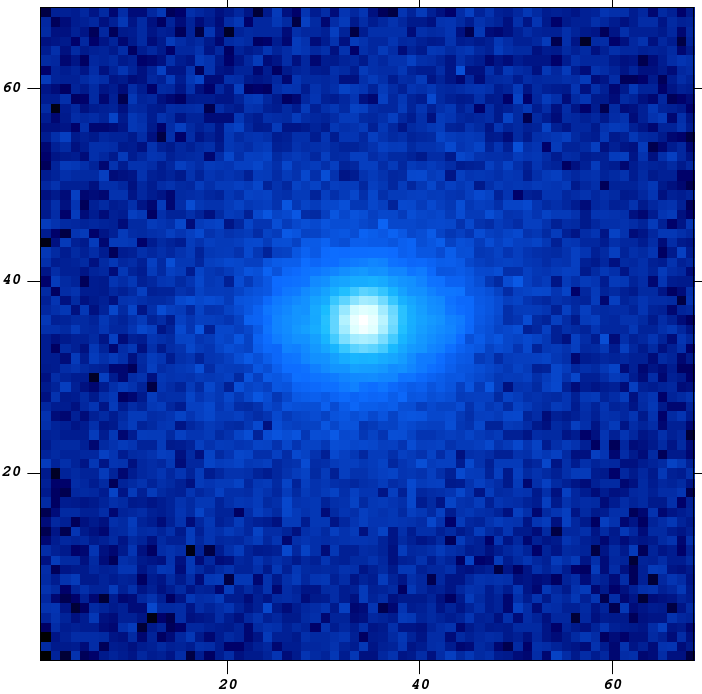}
      \includegraphics[width=1.6cm]{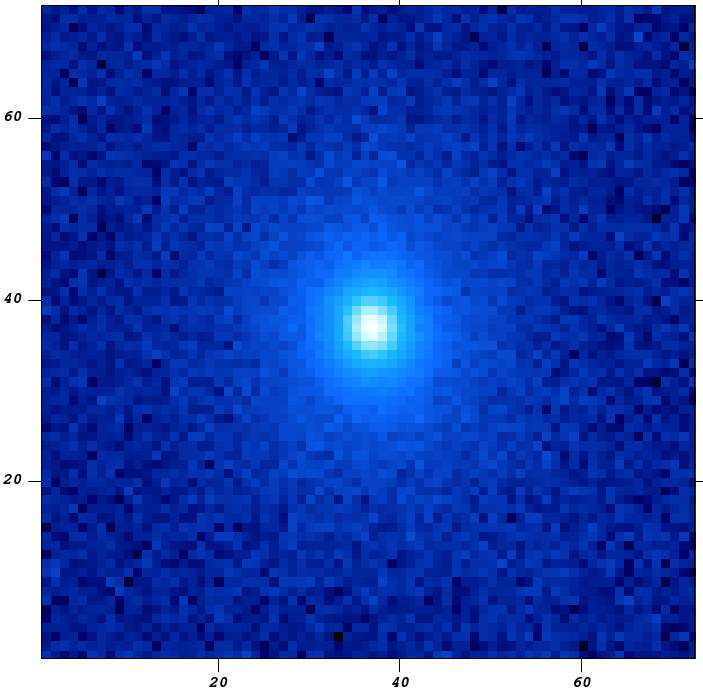}
      \includegraphics[width=1.6cm]{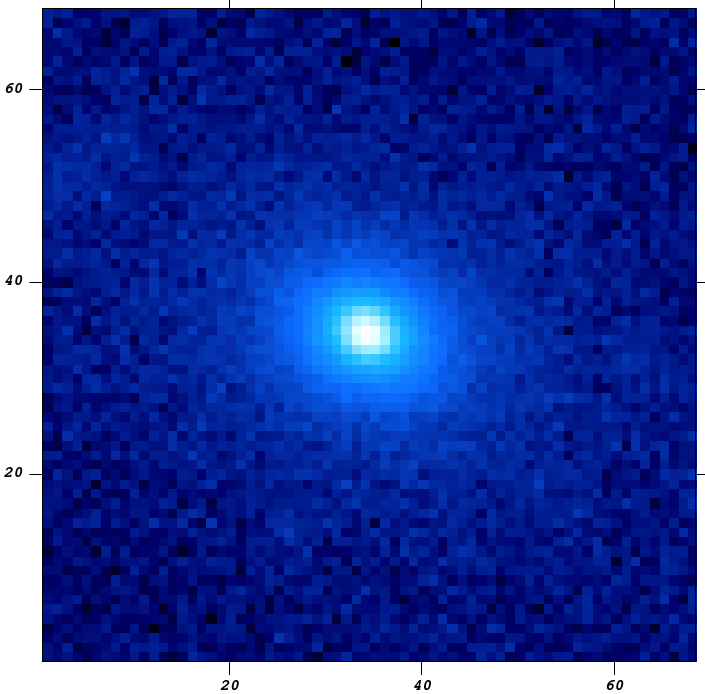}
      \includegraphics[width=1.6cm]{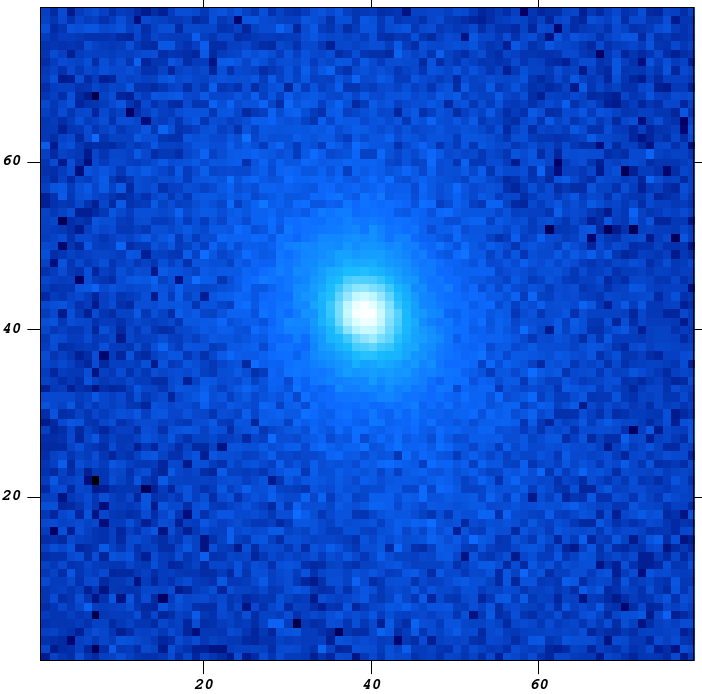}
${}$\\*
      \includegraphics[width=1.6cm]{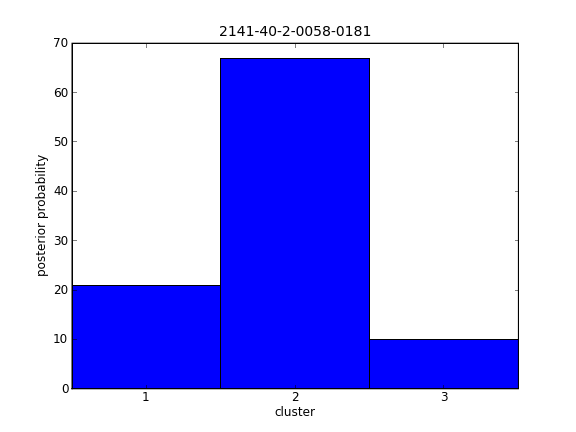}
      \includegraphics[width=1.6cm]{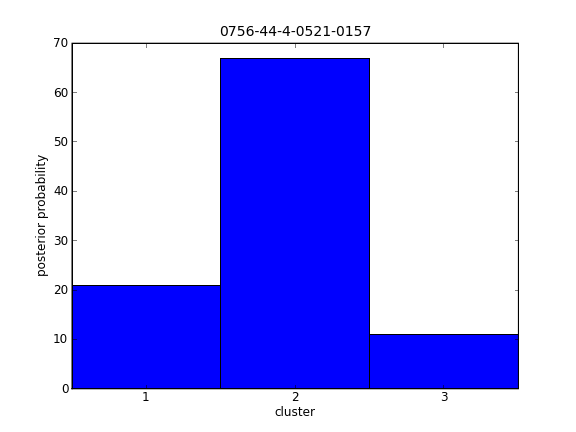}
      \includegraphics[width=1.6cm]{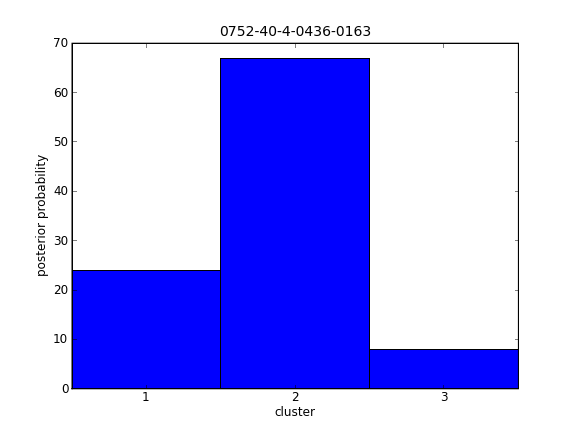}
      \includegraphics[width=1.6cm]{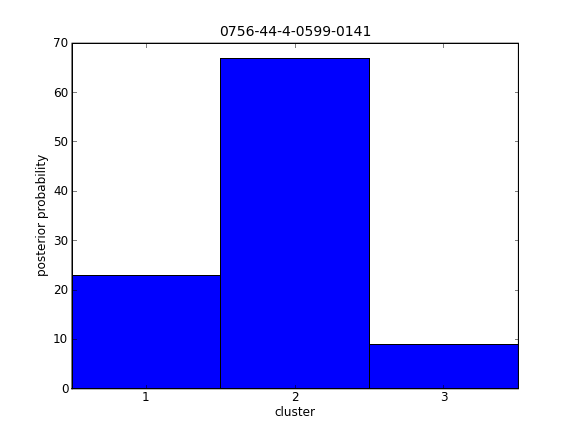}
      \includegraphics[width=1.6cm]{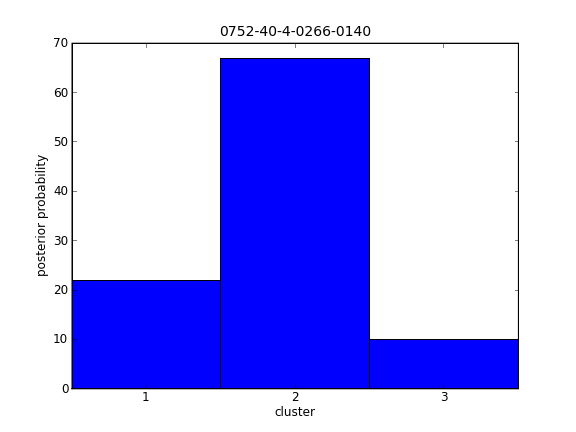}
${}$\\*
      \includegraphics[width=1.6cm]{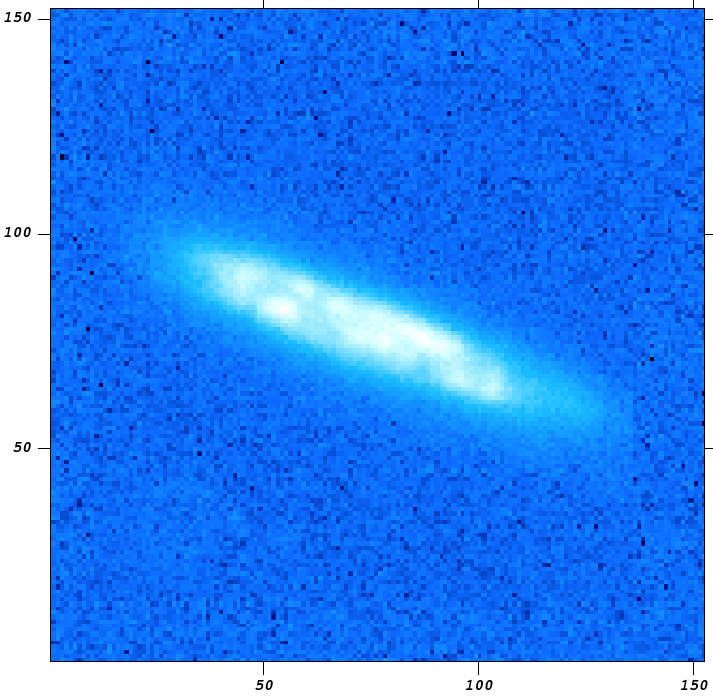}
      \includegraphics[width=1.6cm]{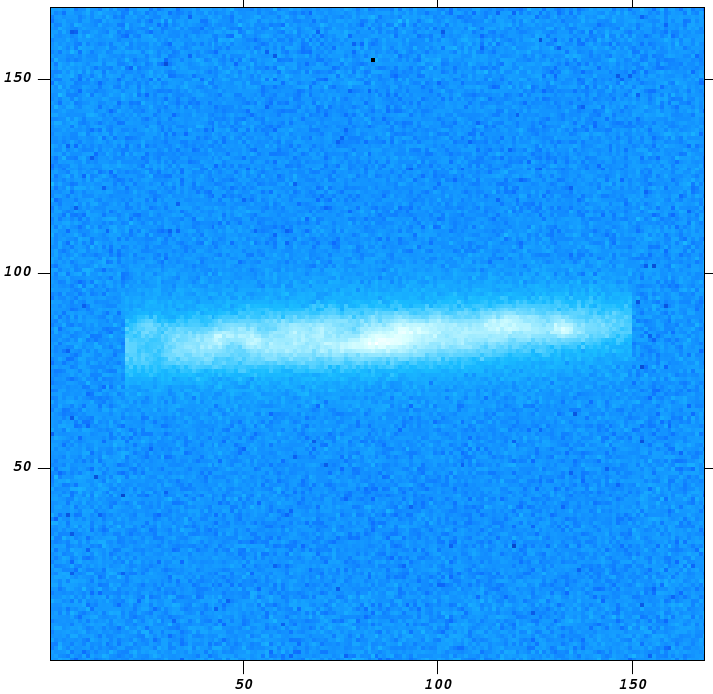}
      \includegraphics[width=1.6cm]{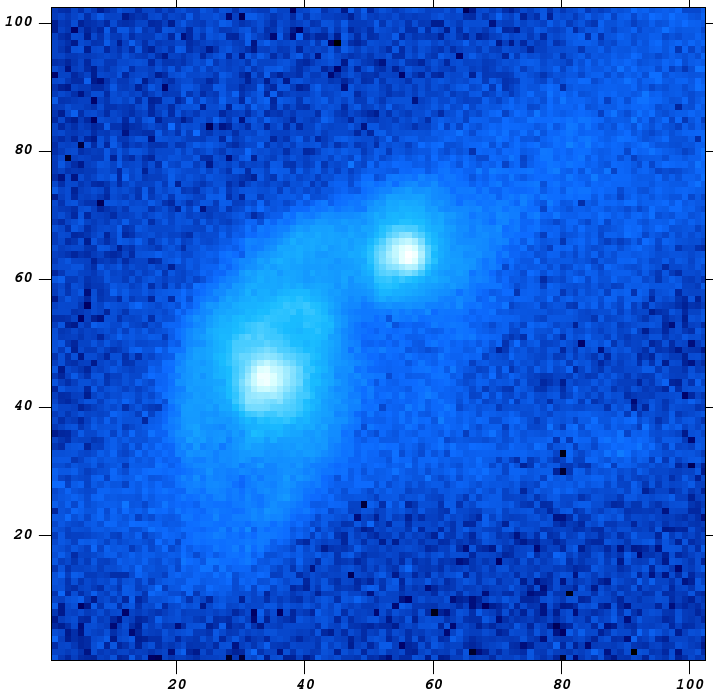}
      \includegraphics[width=1.6cm]{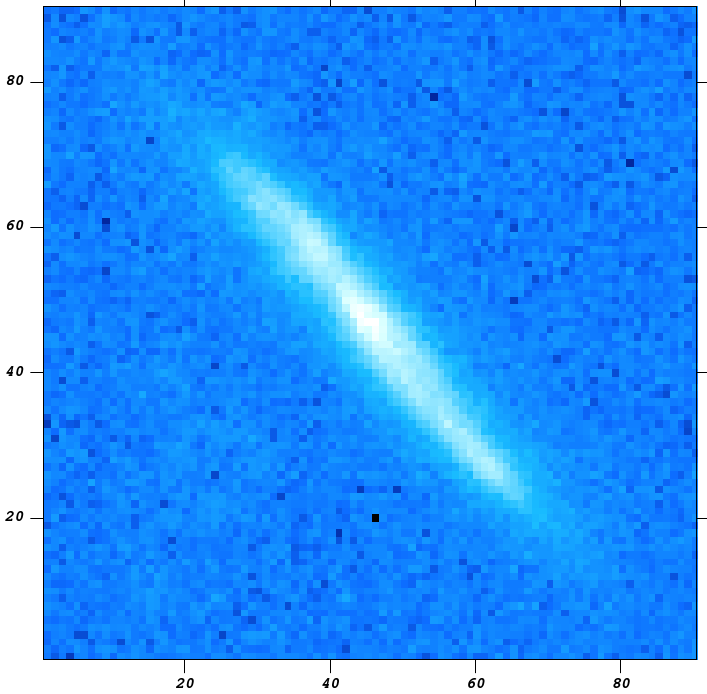}
      \includegraphics[width=1.6cm]{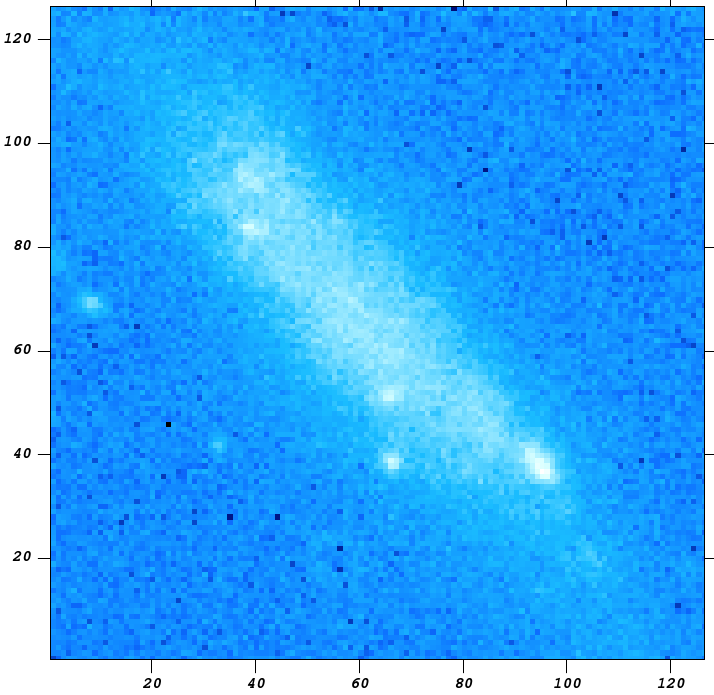}
${}$\\*
      \includegraphics[width=1.6cm]{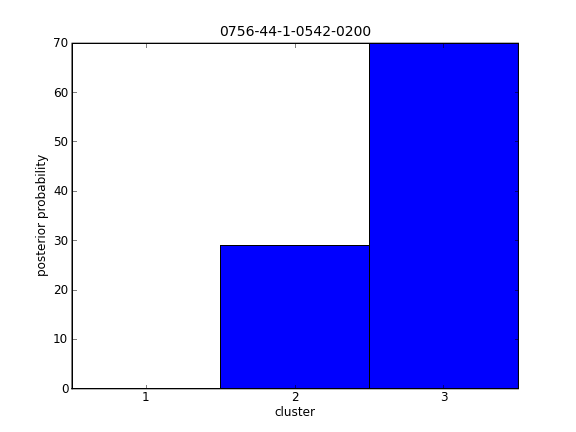}
      \includegraphics[width=1.6cm]{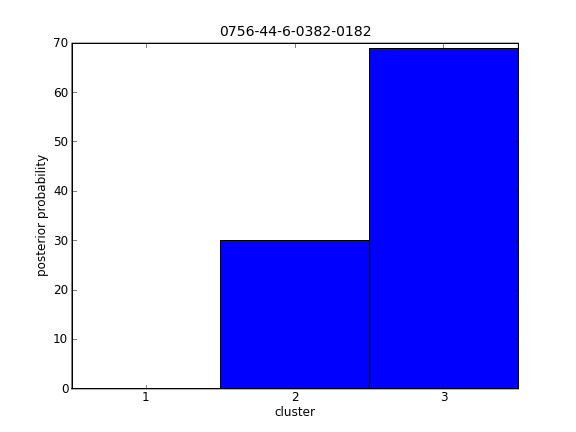}
      \includegraphics[width=1.6cm]{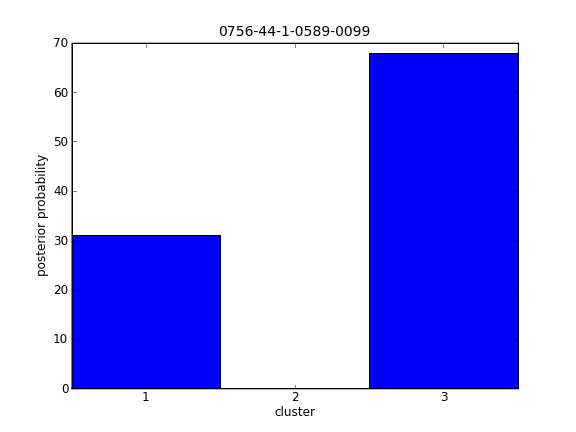}
      \includegraphics[width=1.6cm]{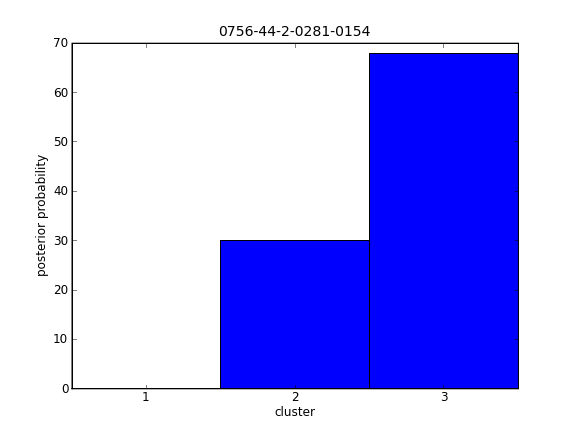}
      \includegraphics[width=1.6cm]{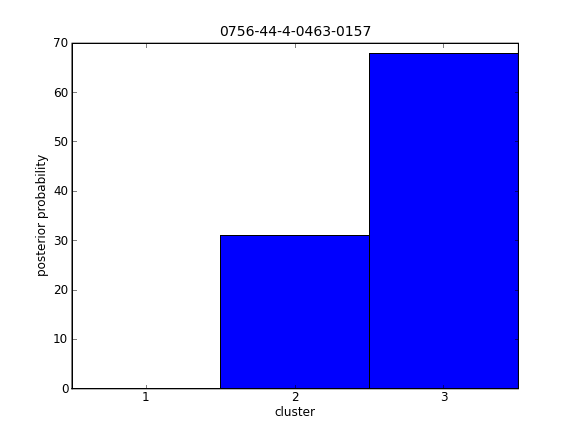}

\caption{Top example objects for $K=3$ clusters.}
Each row corresponds to a cluster. We also show the distribution of its cluster posteriors beneath each object. Cluster 1 seems to contain face-on disks, cluster 2 compact objects, and cluster 3 edge-on disks.
\label{fig:results_Fuku_K3}
\end{figure}

First, we investigate the clustering results for $K=3$, where we have $\SSR\approx 6,146$ (cf. Table \ref{table:fit_results_small_sample}) corresponding to an RMS residual of $\approx 3.7\%$ (cf. Eq. (\ref{eq:mean_residual_from_SSR})) for the similarity-matrix reconstruction. In Fig. \ref{fig:results_Fuku_K3} we show the top five example objects for each of the three clusters together with a histogram of the distribution of cluster posteriors. Inspecting the example images, we clearly see that the first cluster is obviously composed of face-on disk galaxies, whereas the second cluster contains ellipticals. The third cluster is the cluster of edge-on disk galaxies or disks with high inclination angles. However, a blended object has been misclassified into this cluster, too. There are still some blended objects left that we failed to remove, since when sorting out blended objects we visually inspected the images in reduced resolution. The cluster posteriors for $K=3$ are very informative: First, we notice that objects from cluster 1 have typically very low posteriors in cluster 2 and intermediate posteriors in cluster 3, i.e. face-on disks are more similar to edge-on disks than to ellipticals. Second, objects from cluster 2 have low posteriors in all other clusters. Third, objects from cluster 3 tend to be more similar to objects in cluster 2, i.e. edge-on disks are more similar to ellipticals. This is probably due to the higher light concentration and steep light profiles.

\begin{figure}
      \includegraphics[width=8cm]{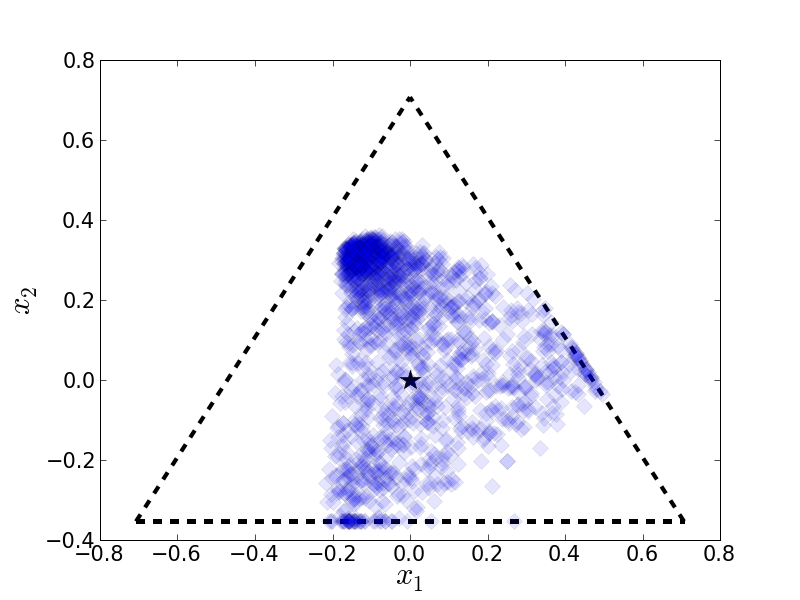}
\caption{Cluster posterior space for $K=3$.}
\label{fig:Fuku_posterior_plane_K3}
Projected cluster posteriors are displayed 10\% translucent such that their density becomes visible. See Fig. \ref{fig:3_class_testbed_posterior_planes} for an explanation of the topology of this plot.
\end{figure}

These results demonstrate that the clustering analysis indeed yields reasonable results for realistic data sets. Furthermore, the results for three clusters are very similar to the clustering scheme of Sect. \ref{sect:testbed_FON_EON_ELL}. However, three clusters are not enough to describe the data faithfully. This is evident from the much larger SSR value for $K=3$ compared to $K=8$ and from Fig. \ref{fig:Fuku_posterior_plane_K3}, where we show the resulting cluster posterior space for $K=3$. Large parts of the available posterior space remain empty whereas the central region is crowded. This behaviour is due to the lack of complexity in the bipartite-graph model and strongly suggests that more clusters are necessary.

\begin{figure*}
  \includegraphics[width=17cm]{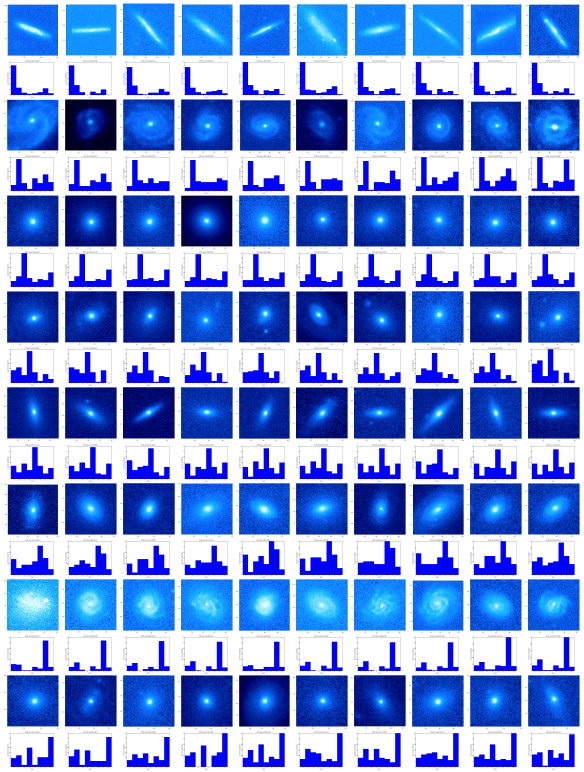}
\caption{Top example objects for $K=8$ clusters.}
Each row corresponds to a cluster. For each object, we also show the histogram of the distribution of its cluster posteriors beneath it. The objects were aligned in shapelet space, not in real space.
\label{fig:results_Fuku_K8}
\end{figure*}

For $K=8$ we have $\SSR\approx 2,254$ (cf. Table \ref{table:fit_results_small_sample}), which corresponds to an RMS residual of $\approx 2.2\%$ for the similarity-matrix reconstruction. We show ten top example objects for each cluster in Fig. \ref{fig:results_Fuku_K8}. First, we notice that the resulting grouping is excellent. However, it is difficult to understand the differences between some clusters. Clusters 1 and 5 are obviously objects with high ellipticities, e.g. edge-on disks, but what is their difference? Is it the bulge dominance which is much weaker in cluster 1 than in 5? Do the clusters differ in their radial light profiles? What is the difference between clusters 2 and 7 which are both face-on disks? Of particular interest are clusters 3 and 8, where both seem to contain roundish and compact objects. However, the posterior histograms reveal a highly asymmetric relation: Objects from cluster 3 also prefer cluster 8 above all other clusters. Nevertheless, most of the top examples of cluster 8 have extremely low posteriors in cluster 3, i.e. association with cluster 3 is highly disfavoured. Although we cannot explain this result without further investigation, it is interesting that the algorithm picked up such a distinctive signal.

\begin{figure}
     \includegraphics[width=8cm]{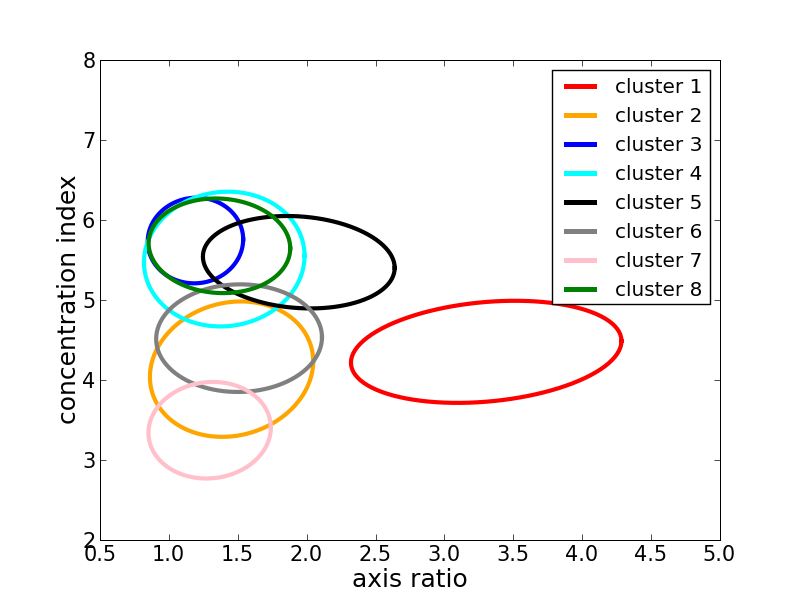}
\caption{Mean axis ratios and concentration indices for the clusters of Fig. \ref{fig:results_Fuku_K8}.}
Weighted means were computed from the top 100 example objects of each cluster. We show the contours of $1\sigma$ and take into account possible correlations.
\label{fig:Fuku_cluster_axis_vs_concentration}
\end{figure}

As we have access to the isophotal axis ratio and the concentration index (cf. Eq. (\ref{eq:concentration_index})) for all objects, we investigate their distributions for the clusters. Figure \ref{fig:Fuku_cluster_axis_vs_concentration} shows the mean axis ratios and the mean concentration indices for all eight clusters averaged over the 100 top examples. The cluster with the highest mean axis ratio is cluster 1, which is the cluster of edge-on disk galaxies. The cluster with lowest concentration index is cluster 7, which is the cluster of face-on disk galaxies that exhibit extended smooth light profiles. Clusters 3, 4, 5 and 8 have the largest concentration indices. As is evident from Fig. \ref{fig:results_Fuku_K8}, these clusters are indeed composed of rather compact objects that seem to be elliptical galaxies. However, there is no decisive distinction in Fig. \ref{fig:Fuku_cluster_axis_vs_concentration}. This is not necessarily a flaw in the clustering results, but rather more likely caused by concentration and axis ratio being an insufficient parametrisation scheme \citep[cf.][]{Andrae2010b}.

It seems like  the resulting classification scheme is essentially face-on disk, edge-on disk and elliptical. If we increase the number of clusters, we get further diversification that may be caused by bulge dominance or inclination angles. We emphasise again that our primary goal is to demonstrate that our method discovers morphological classes and provides data-to-class assignments that are reasonable.

\section{Conclusions\label{sect:conclusions}}

We briefly summarise our most important arguments and results:
\begin{itemize}
\item Galaxy evolution, the process of observation, and the experience with previous classification attempts strongly suggest a probabilistic (''soft'') classification. Hard classifications appear to be generically inappropriate.
\item There are two distance-based soft-clustering algorithms that have been applied to galaxy morphologies so far: Gaussian mixture models by \citet{Kelly2004,Kelly2005} and the bipartite-graph model by \citet{Yu2005} presented in this work. The weak points of the Gaussian mixture model are the dimensionality reduction and its assumption of Gaussianity. The weakness of the bipartite-graph model is the definition of the similarity measure.
\item The shapelet formalism, our similarity measure, and the bipartite-graph model produce reasonable clusters and data-to-cluster assignments for real galaxies. The automated discovery of classes corresponding to face-on disks, edge-on disks and elliptical galaxies without any prior assumptions is impressive and demonstrates the great potential of clustering analysis. Moreover, the automatically discovered classes have a qualitatively different meaning compared to pre-defined classes, since they represent to grouping that is preferred by the given data sample itself.
\item Random effects such as orientation angle and inclination are a major obstacle, since they introduce additional scatter into a parametrisation of galaxy morphologies.
\item For data sets containing $N$ galaxies, the computation times scale as $\mathcal O(N^2)$. Nevertheless, we experienced that a clustering analysis is feasible for data sets containing up to $N=10,000$ galaxies \textit{without} employing supercomputers. We conclude that a clustering analysis on a data set of one million galaxies is possible using supercomputers.
\item It is possible to enhance this method by setting up a classifier based on the classes found by the soft-clustering analysis, thereby improving the time complexity from $\mathcal O(N^2)$ to $\mathcal O(N)$.
\item The method presented in this paper is not limited to galaxy morphologies only. For instances, it could possibly be applied to automated star-galaxy classification or AGN detection.
\end{itemize}
The bottom line of this paper is that automatic discovery of morphological classes and object-to-class assignments (clustering analysis) does work and is less prejudiced and time-consuming than visual classifications, though the interpretation of the results is still an open issue. Especially when analysing new data samples for the first time, clustering algorithms are more objective than using pre-defined classes and visual classifications. The advantages of such a sophisticated statistical algorithm justify its considerable complexity.

\begin{acknowledgements}
RA thanks Coryn Bailer-Jones for his comments that have been particularly useful to improve this work. Furthermore, RA thanks Knud Jahnke and Eric Bell for helpful discussions especially on the interpretation of the results. RA is funded by a Klaus-Tschira scholarship. PM is supported by the DFG Priority Programme 1177.
\end{acknowledgements}

\bibliographystyle{aa}

\begin{thebibliography}{20}
\expandafter\ifx\csname natexlab\endcsname\relax\def\natexlab#1{#1}\fi

\bibitem[{{Abazajian} {et~al.}(2005){Abazajian}, {Adelman-McCarthy},
  {Ag{\"u}eros}, {Allam}, {Anderson}, {Anderson}, {Annis}, {Bahcall}, {Baldry},
  {Bastian}, {Berlind}, {Bernardi}, {Blanton}, {Bochanski}, {Boroski},
  {Brewington}, {Briggs}, {Brinkmann}, {Brunner}, {Budav{\'a}ri}, {Carey},
  {Castander}, {Connolly}, {Covey}, {Csabai}, {Dalcanton}, {Doi}, {Dong},
  {Eisenstein}, {Evans}, {Fan}, {Finkbeiner}, {Friedman}, {Frieman},
  {Fukugita}, {Gillespie}, {Glazebrook}, {Gray}, {Grebel}, {Gunn}, {Gurbani},
  {Hall}, {Hamabe}, {Harbeck}, {Harris}, {Harris}, {Harvanek}, {Hawley},
  {Hayes}, {Heckman}, {Hendry}, {Hennessy}, {Hindsley}, {Hogan}, {Hogg},
  {Holmgren}, {Holtzman}, {Ichikawa}, {Ichikawa}, {Ivezi{\'c}}, {Jester},
  {Johnston}, {Jorgensen}, {Juri{\'c}}, {Kent}, {Kleinman}, {Knapp}, {Kniazev},
  {Kron}, {Krzesinski}, {Lamb}, {Lampeitl}, {Lee}, {Lin}, {Long}, {Loveday},
  {Lupton}, {Mannery}, {Margon}, {Mart{\'{\i}}nez-Delgado}, {Matsubara},
  {McGehee}, {McKay}, {Meiksin}, {M{\'e}nard}, {Munn}, {Nash}, {Neilsen},
  {Newberg}, {Newman}, {Nichol}, {Nicinski}, {Nieto-Santisteban}, {Nitta},
  {Okamura}, {O'Mullane}, {Owen}, {Padmanabhan}, {Pauls}, {Peoples}, {Pier},
  {Pope}, {Pourbaix}, {Quinn}, {Raddick}, {Richards}, {Richmond}, {Rix},
  {Rockosi}, {Schlegel}, {Schneider}, {Schroeder}, {Scranton}, {Sekiguchi},
  {Sheldon}, {Shimasaku}, {Silvestri}, {Smith}, {Smol{\v c}i{\'c}}, {Snedden},
  {Stebbins}, {Stoughton}, {Strauss}, {SubbaRao}, {Szalay}, {Szapudi},
  {Szkody}, {Szokoly}, {Tegmark}, {Teodoro}, {Thakar}, {Tremonti}, {Tucker},
  {Uomoto}, {Vanden Berk}, {Vandenberg}, {Vogeley}, {Voges}, {Vogt},
  {Walkowicz}, {Wang}, {Weinberg}, {West}, {White}, {Wilhite}, {Xu}, {Yanny},
  {Yasuda}, {Yip}, {Yocum}, {York}, {Zehavi}, {Zibetti}, \&
  {Zucker}}]{Abazajian2005}
{Abazajian}, K., {Adelman-McCarthy}, J.~K., {Ag{\"u}eros}, M.~A., {et~al.}
  2005, \aj, 129, 1755

\bibitem[{{Andrae} {et~al.}(in prep.){Andrae}, {Melchior}, \&
  {Jahnke}}]{Andrae2010b}
{Andrae}, R., {Melchior}, P., \& {Jahnke}, K. in prep.

\bibitem[{{Baldry} {et~al.}(2004){Baldry}, {Glazebrook}, {Brinkmann},
  {Ivezi{\'c}}, {Lupton}, {Nichol}, \& {Szalay}}]{Baldry2004}
{Baldry}, I.~K., {Glazebrook}, K., {Brinkmann}, J., {et~al.} 2004, \apj, 600,
  681

\bibitem[{{Bamford} {et~al.}(2009){Bamford}, {Nichol}, {Baldry}, {Land},
  {Lintott}, {Schawinski}, {Slosar}, {Szalay}, {Thomas}, {Torki}, {Andreescu},
  {Edmondson}, {Miller}, {Murray}, {Raddick}, \& {Vandenberg}}]{Bamford2009}
{Bamford}, S.~P., {Nichol}, R.~C., {Baldry}, I.~K., {et~al.} 2009, \mnras, 393,
  1324

\bibitem[{Bellman(1961)}]{Bellman1961}
Bellman, R. 1961, Adaptive Control Processes: A Guided Tour (Princeton
  University Press)

\bibitem[{Bilmes(1997)}]{Bilmes1997}
Bilmes, J.~A. 1997, International Computer Science Institute, technical report
  TR-97-021

\bibitem[{{Conselice}(2003)}]{Conselice2003}
{Conselice}, C.~J. 2003, {The Relationship between Stellar Light Distributions
  of Galaxies and Their Formation Histories}

\bibitem[{{Croton} {et~al.}(2006){Croton}, {Springel}, {White}, {De Lucia},
  {Frenk}, {Gao}, {Jenkins}, {Kauffmann}, {Navarro}, \& {Yoshida}}]{Croton2006}
{Croton}, D.~J., {Springel}, V., {White}, S.~D.~M., {et~al.} 2006, \mnras, 365,
  11

\bibitem[{{Fukugita} {et~al.}(2007){Fukugita}, {Nakamura}, {Okamura}, {Yasuda},
  {Barentine}, {Brinkmann}, {Gunn}, {Harvanek}, {Ichikawa}, {Lupton},
  {Schneider}, {Strauss}, \& {York}}]{Fukugita2007}
{Fukugita}, M., {Nakamura}, O., {Okamura}, S., {et~al.} 2007, \aj, 134, 579

\bibitem[{{Kelly} \& {McKay}(2004)}]{Kelly2004}
{Kelly}, B.~C. \& {McKay}, T.~A. 2004, \aj, 127, 625

\bibitem[{{Kelly} \& {McKay}(2005)}]{Kelly2005}
{Kelly}, B.~C. \& {McKay}, T.~A. 2005, \aj, 129, 1287

\bibitem[{{Massey} \& {R\'efr\'egier}(2005)}]{Massey2005}
{Massey}, R. \& {R\'efr\'egier}, A. 2005, \mnras, 363, 197

\bibitem[{{Melchior} {et~al.}(2009{\natexlab{a}}){Melchior}, {Andrae},
  {Maturi}, \& {Bartelmann}}]{Melchior2009}
{Melchior}, P., {Andrae}, R., {Maturi}, M., \& {Bartelmann}, M.
  2009{\natexlab{a}}, \aap, 493, 727

\bibitem[{{Melchior} {et~al.}(2009{\natexlab{b}}){Melchior}, {Boehnert},
  {Lombardi}, \& {Bartelmann}}]{Melchior2009a}
{Melchior}, P., {Boehnert}, A., {Lombardi}, M., \& {Bartelmann}, M.
  2009{\natexlab{b}}, ArXiv e-prints

\bibitem[{{Melchior} {et~al.}(2007){Melchior}, {Meneghetti}, \&
  {Bartelmann}}]{Melchior2007}
{Melchior}, P., {Meneghetti}, M., \& {Bartelmann}, M. 2007, \aap, 463, 1215

\bibitem[{Redner \& Walker(1984)}]{Redner1984}
Redner, R. \& Walker, H. 1984, SIAM Review, 26, 195

\bibitem[{{R\'efr\'egier}(2003)}]{Refregier2003}
{R\'efr\'egier}, A. 2003, \mnras, 338, 35

\bibitem[{{Richards} {et~al.}(2009){Richards}, {Freeman}, {Lee}, \&
  {Schafer}}]{Richards2009}
{Richards}, J.~W., {Freeman}, P.~E., {Lee}, A.~B., \& {Schafer}, C.~M. 2009,
  \apj, 691, 32

\bibitem[{{Strateva} {et~al.}(2001){Strateva}, {Ivezi{\'c}}, {Knapp},
  {Narayanan}, {Strauss}, {Gunn}, {Lupton}, {Schlegel}, {Bahcall}, {Brinkmann},
  {Brunner}, {Budav{\'a}ri}, {Csabai}, {Castander}, {Doi}, {Fukugita}, {Gy{\H
  o}ry}, {Hamabe}, {Hennessy}, {Ichikawa}, {Kunszt}, {Lamb}, {McKay},
  {Okamura}, {Racusin}, {Sekiguchi}, {Schneider}, {Shimasaku}, \&
  {York}}]{Strateva2001}
{Strateva}, I., {Ivezi{\'c}}, {\v Z}., {Knapp}, G.~R., {et~al.} 2001, \aj, 122,
  1861

\bibitem[{{Yu} {et~al.}(2005){Yu}, {Yu}, \& {Tresp}}]{Yu2005}
{Yu}, K., {Yu}, S., \& {Tresp}, V. 2005, Advances in Neural Information
  Processing Systems

\end{thebibliography}

\def\physrep{Phys. Rep.}%
\def\apjs{ApJS}%
\def\apj{ApJ}%
\def\aj{AJ}%
\def\aap{A\&A}%
\def\aaps{A\&AS}%
\def\mnras{MNRAS}%

\end{document}